\documentclass[10pt, conference]{IEEEtran}

\IEEEoverridecommandlockouts
\usepackage{booktabs}
\usepackage[utf8]{inputenc}
\usepackage{amsmath}
\usepackage{amsfonts}
\usepackage{amssymb}
\usepackage{graphicx}
\usepackage{listings}
\usepackage{algorithm}
\usepackage[noend]{algpseudocode}
\usepackage[utf8]{inputenc}
\usepackage[english]{babel}
\usepackage{xspace}
\usepackage{tabularx}
\usepackage{paralist}
\usepackage{multirow}
\usepackage[table,xcdraw]{xcolor}
\usepackage[T1]{fontenc}
\usepackage[scaled=.8]{beramono}
\def\BibTeX{{\rm B\kern-.05em{\sc i\kern-.025em b}\kern-.08em
    T\kern-.1667em\lower.7ex\hbox{E}\kern-.125emX}}

\newcommand{\tool}{\textsc{CoPro}\xspace}
\newcolumntype{L}[1]{>{\raggedright\arraybackslash}p{#1}}

\usepackage{amsthm}

\usepackage{balance}
\usepackage{xcolor}
\let\labelindent\relax
\usepackage{enumitem}

\newtheorem{definition}{Definition}

\newcommand{\code}[1]{{\small\textsf{#1}}}

\newcommand{\eg}{\hbox{\emph{e.g.,}}\xspace}
\newcommand{\ie}{\hbox{\emph{i.e.,}}\xspace}
\newcommand{\etc}{\hbox{\emph{etc.}}\xspace}

\definecolor{dkgreen}{rgb}{0,0.6,0}
\definecolor{gray}{rgb}{0.5,0.5,0.5}
\definecolor{mauve}{rgb}{0.58,0,0.82}
\makeatletter
\renewcommand{\ALG@beginalgorithmic}{\small}
\makeatother

\ifCLASSINFOpdf
\else
\fi

\begin{document}
%

\title{Feature-Interaction Aware Configuration Prioritization for Configurable Code}


\author{\IEEEauthorblockN{Son Nguyen\IEEEauthorrefmark{1},
Hoan Nguyen\IEEEauthorrefmark{2},
Ngoc Tran\IEEEauthorrefmark{1},
Hieu Tran\IEEEauthorrefmark{1}, and
Tien N. Nguyen\IEEEauthorrefmark{1}}
  \IEEEauthorblockA{\IEEEauthorrefmark{1}Computer Science Department, The University of Texas at Dallas, USA,\\ \IEEEauthorrefmark{1}Email: \{sonnguyen,nmt140230,trunghieu.tran,tien.n.nguyen\}@utdallas.edu}
  \IEEEauthorblockA{\IEEEauthorrefmark{2}Amazon Corporation, USA, Email: nguyenanhhoan@gmail.com}}



%


\maketitle



\begin{abstract}
Unexpected interactions among features induce most bugs in a
configurable software system. Exhaustively analyzing all the
exponential number of possible configurations is prohibitively costly.
Thus, various sampling techniques have been proposed to systematically
narrow down the exponential number of legal configurations to be
analyzed.  Since analyzing all selected configurations can require a
huge amount of effort, fault-based configuration prioritization, that
helps detect faults earlier, can yield practical benefits in quality
assurance. In this paper, we propose {\tool}, a novel formulation of
feature-interaction bugs via common program entities enabled/disabled
by the features. Leveraging from that, we develop an efficient
feature-interaction-aware configuration prioritization technique for a
configurable system by ranking the configurations according to their
total number of potential bugs. We conducted several experiments to
evaluate {\tool} on the ability to detect configuration-related bugs
in a public benchmark. We found that {\tool} outperforms the
state-of-the-art configuration prioritization techniques when we add
them on advanced sampling algorithms. In 78\% of the cases, {\tool}
ranks the buggy configurations at the top 3 positions in the resulting
list. Interestingly, {\tool} is able to detect 17 not-yet-discovered
feature-interaction bugs.
\end{abstract}


%
\IEEEpeerreviewmaketitle

\section{Introduction}
\label{sec:intro}

Several software systems enable developers to configure to different
environments and requirements. In practice, a highly-configurable
system can tailor its functional and non-functional properties to the
needs and requirements of users. It does so via 
a very large number of \textit{configuration
  options}~\cite{6572787,Berger:2013:SVM:2430502.2430513} that are
used to control different
\textit{features}~\cite{apel8overview,kang1990feature}.
For example, Linux Kernel supports more than 12,000 compile-time
configuration options, that can be configured to generate specific
kernel \textit{variants} for billions of~scenarios.

In a configurable system, features can interact with one another in a
non-trivial manner. As a consequence, such interaction could
inadvertently modify or influence the functionality of one
another~\cite{Zave:2003:EFE:766951.766969}. Unexpected interactions
might induce bugs. In fact, most configuration-related bugs are
caused by interactions among
features~\cite{Abal:2014:VBL:2642937.2642990,Garvin:2011:FIF:2120102.2120918,Medeiros:2016:CSA:2884781.2884793,Nie:2011:SCT:1883612.1883618,Thum:2014:CSA:2620784.2580950}. Unfortunately,
traditional methods cannot be directly applied to work on configurable
code since they focus on detecting bugs in a particular variant.
%
Furthermore, exhaustively analyzing the systems is infeasible due to
the exponential number of all possible configurations.
In practice, configuration testing is often performed in a manual
and ad-hoc manner by unsystematically selecting common variants for
analysis~\cite{Greiler:2012:TCS:2337223.2337253,Machado:2014:STS:2658281.2658306}. 


%

\begin{figure}[t]
\centering
\includegraphics[width=3.4in]{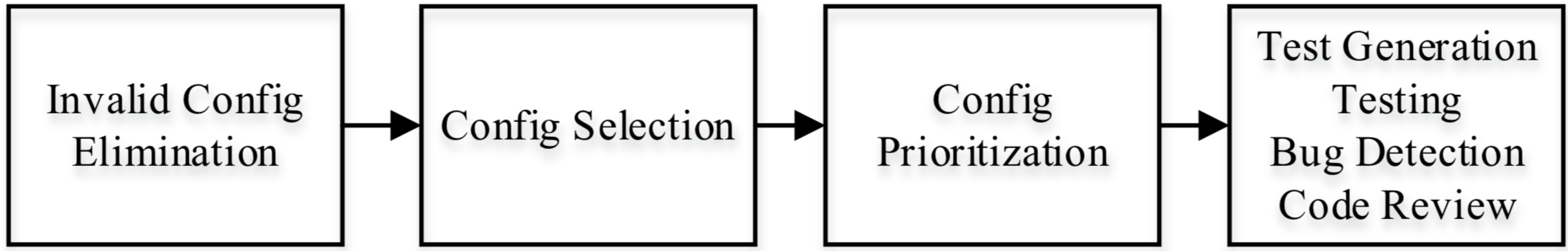}
\caption{The QA Process of Configurable System}
\label{process}
\end{figure}

To systematically perform quality assurance (QA) for a
highly-configurable system (Figure \ref{process}), researchers have
proposed several techniques to narrow the configuration space {\em by
  {\bf eliminating invalid configurations} that violate the feature
  model} of the system, which defines the feasible configurations via
the constraints among the
features~\cite{Classen:2011:SMC:1985793.1985838,Classen:2010:MCL:1806799.1806850,Gruler:2008:MMC:1424547.1424557,kastner2012virtual,kang1990feature,
  Post:2008:CLV:1642931.1642971}. However,
the number of configurations that need to be tested is still
exponential. To address this explosion problem, researchers introduce
various {\bf configuration selection} strategies. The popular
strategies include the sampling algorithms which achieve feature
interaction coverage
such as \textit{combinatorial
  interaction testing}~\cite{Perrouin:2010:AST:1828417.1828490,Oster:2010:AIP:1885639.1885658,johansen-splc12,Marijan:2013:PPT:2491627.2491646}, \textit{one-enabled}~\cite{Medeiros:2016:CSA:2884781.2884793}, \textit{one-disabled}~\cite{Abal:2014:VBL:2642937.2642990}, \textit{most-enabled-disabled}~\cite{tartler2014static},
\textit{statement-coverage}~\cite{Tartler:2012:CCA:2094091.2094095},
to reduce the number of configurations to be analyzed.
Still, \textit{those algorithms assume the chances of detecting
  interaction bugs are the same for all those combinations}. Thus,
interaction faults might be discovered only after the last variants in
such samples is tested. Thus, after configuration selection, 
the selected {\em set of configurations need to be {\bf prioritized}}
for QA activities~\cite{Al-Hajjaji:2014:SPS:2648511.2648532}. Note
that configuration prioritization is different from test case
prioritization because after configuration prioritization, any QA
activities can be applied on the ranked list of prioritized
configurations, including test generation and testing, static bug detection, or manual
code review (Figure~\ref{process}).

To motivate configuration prioritization, let us take an example of
Linux Kernel. In Linux, the number of different configuration options
is over 12,000, leading to +$2^{12K}$ different configurations. After
applying all the constraints on various combinations of options, the
number of valid configurations for QA is an exponential number.
For configuration selection, by using \textit{six-wise} sampling
algorithm, the number is still extremely large, up to +500K
configurations~\cite{Medeiros:2016:CSA:2884781.2884793}.
Hence, without configuration prioritization, many 
bugs that are dependent on configurations might still be hidden
due to this large configuration space,
especially when the resources for QA (\eg time and developers'
efforts) are limited.

In practice, developers even do not perform QA activities on a
particular configuration until it was reported to have defects by the
users. In this case, users have already suffered the consequences of
those defects.
Due to the large number of configurations after selection for QA, even
{\em compile-time errors} and flaws cannot be quickly detected by a
compiler or a bug detection in the appropriate configuration. Indeed,
in the Variability Bugs Database
(VBDb)~\cite{Abal:2014:VBL:2642937.2642990}, a public database of
real-world configuration-related bugs reported for the Linux kernel,
{\em there are 42 out of 98 bugs and flaws that are compile-time: 25
  declaration errors, 10 type errors, and 7 cases of dead~code}.

Despite the importance of configuration prioritization, the
state-of-the-art methods for such prioritization are still limited in
detecting feature-interaction bugs.
The similarity-based prioritization method
($SP$)~\cite{Al-Hajjaji:2014:SPS:2648511.2648532} is based on the idea
that~dissimilar test sets are likely to detect more defects than
similar ones~\cite{Al-Hajjaji:2014:SPS:2648511.2648532}. In $SP$, the
configuration with the maximum number of features is selected to be
the first one under test.
%
The next configuration under test is the configuration with the lowest
feature similarity compared to the previously chosen one.
Despite its success, there are two key problems with $SP$. First, $SP$
aims to cover as many features different from the previous ones. The
different features to be considered next might not be the ones that
potentially causes violations. $SP$ does not examine the interaction
between features, which is the key aspect causing interaction bugs in
a variant. Second, in $SP$, the quality of the resulted prioritization
order strongly depends on the selection of the first configuration.

In this paper, we propose {\tool}, a novel configuration
prioritization approach for configurable systems by analyzing their
code to detect feature-interaction bugs. Our key idea in {\tool} is as
follows.
In a configurable system, features are implemented as blocks of code,
which are expressed via the {\em program instructions/operations
  (e.g., declarations, references, assignments, etc.) on the data
  structures/program entities (e.g., variables, functions, etc.)}.
Features interaction occurs when the operations on the program
entities shared between the features have impacts on each other.

{\em Those operations, when the features are enabled or disabled,
  potentially create a violation(s) that makes the program
  not-compilable or having a run-time error}. Detecting feature
interactions via operations would help identify potential
feature-interaction bugs.
%
An example of a violation is that a feature disables the only
initialization of a variable while another enables one of its
dereferences (the violation of {\em ``dereferencing an un-initialized
  variable''}). This violation could lead to a \texttt{NULL} pointer
exception.
It is clear that the configuration in which the former feature is
disabled and the latter is enabled, is more suspicious than the one
where~both of them are either enabled or disabled.
%
The suspiciousness of a configuration is indicated via the potential
feature-interaction violations.
Hence, a higher number of potential violations makes the configuration more suspicious.
The suspiciousness levels are used to rank the configurations,
which helps testing, bug detection, or other QA activities more
efficient.

We conducted several experiments to evaluate {\tool} in two
complementary settings. First, in a benchmark setting, we ran {\tool}
on the Variability Bugs Database
(VBDb)~\cite{Abal:2014:VBL:2642937.2642990}. We compared {\tool} with the
two state-of-the-art approaches in {\em random prioritization} and
{\em similarity-based prioritization}
(SP)~\cite{Al-Hajjaji:2014:SPS:2648511.2648532}, when we added each of
the compared techniques on top of several state-of-the-art sampling
algorithms~\cite{Medeiros:2016:CSA:2884781.2884793}. We found that
{\tool} significantly outperforms the other techniques. In 78.0\% of the
cases, {\tool} ranks the buggy configurations at the top-3 positions
in the list, while the $SP$ and $Random$ approaches rank them at the
top-3 positions for only 41.3\% and 26.1\% of the cases. Interestingly,
{\tool} was able to detect {\em 17 feature-interaction bugs that were not
yet discovered in VBDb} including high-degree interaction bugs, memory
leaking bugs, \etc ~In~the second setting, we
connect {\tool} with a compiler to run on
large, open-source configurable systems, and {\tool}
can detect 4 newly discovered bugs and programming flaws.

In summary, in this paper, our main contributions include:

\begin{compactitem}


\item A formulation of feature-interaction bugs using common
  program entities enabled/disabled by the features;

\item {\tool}: an efficient feature-interaction-aware configuration
  prioritization technique for configurable systems;

\item An extensive experimental evaluation showing the effectiveness
  of {\tool} over the state-of-the-art approaches.

\end{compactitem}

\section{Motivating Example}

\begin{figure}[t]
\centering
\includegraphics[width=0.43\textwidth]{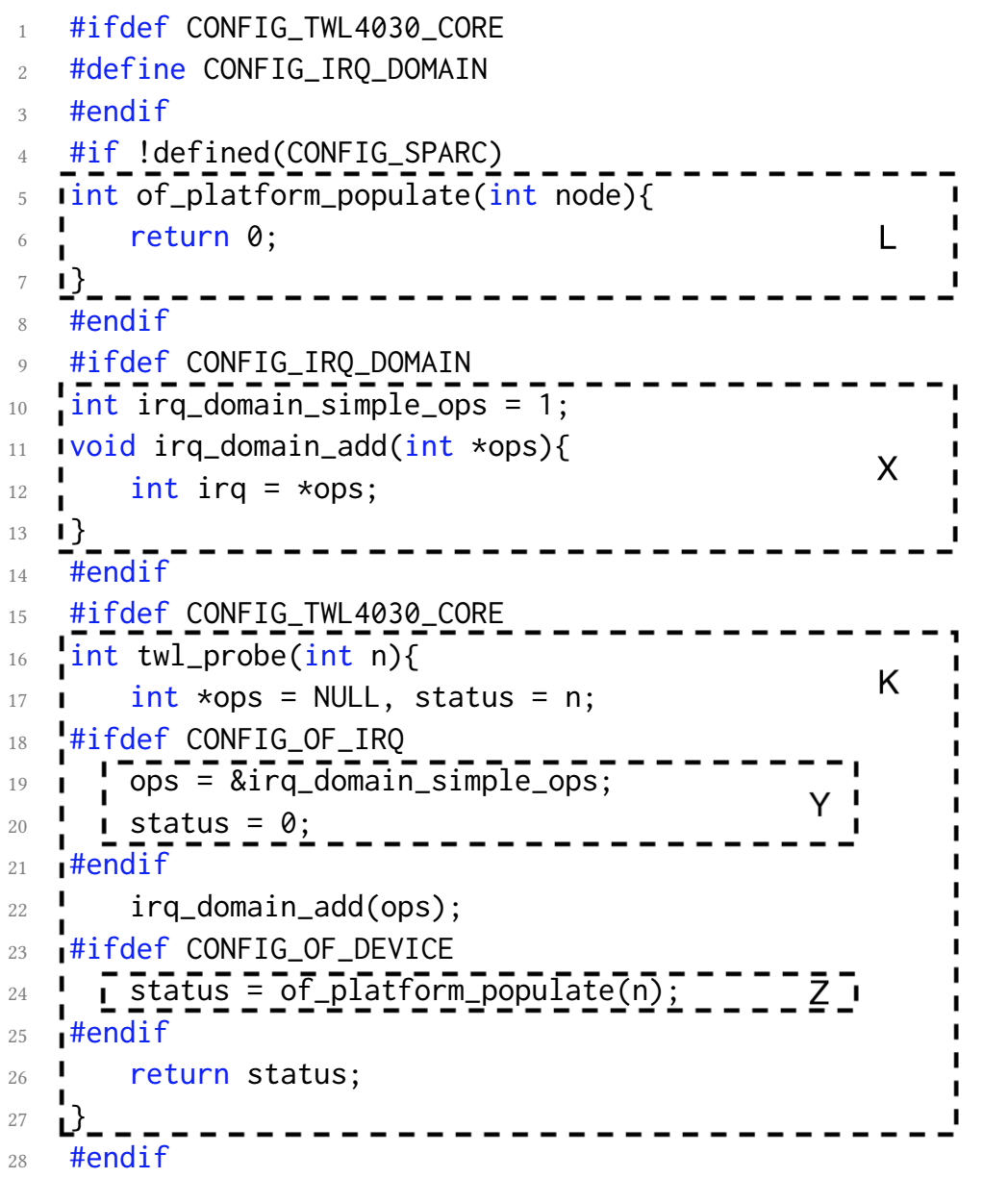} 
\caption{A Simplified Bug in Linux Kernel}
\label{example1}
\end{figure}


In this section, we illustrate the challenges of configuration
prioritization and motivate our solution via an example.

\subsection{Examples of Bugs in Configurable Code}

Let us consider the simplified version of the real buggy code in the
Linux kernel~\cite{Abal:2014:VBL:2642937.2642990} at the commit
40410715715~of Linux-stable at https://git.kernel.org (shown in
Figure~\ref{example1}). This version has more than 5,200 compile-time
configuration options and about 30,000 files.
The code in Figure~\ref{example1} contains two feature-interaction
bugs that were discovered in certain~configurations:



\begin{compactitem}[\textbullet]

\item A compile-time error occurs (\textit{use the undeclared
  function} \texttt{of\_platform\_populate} on line~24) in
  the variants where \texttt{CONFIG\_TWL4030\-\_CORE,
    CONFIG\_OF\_DEVICE, and CONFIG\_SPARC} are enabled.

\item A run-time error occurs (\textit{dereferencing the \texttt{NULL}
  pointer} \texttt{ops} on line~12) in the configurations where
  \texttt{CONFIG\_TWL4030\-\_CORE, CON\-FIG\_OF\_DEVICE} are enabled,
  and \texttt{CONFIG\_SPARC} and \texttt{CONFIG\_OF\_IRQ} are disabled.

\end{compactitem}



For this example in Linux kernel, brute-force testing of all possible
variants to discover these interaction bugs faces the
problem~of~combinatorial explosion in the exponential configuration
space (up to $2^{5,200}$ possible configurations). With a huge number
of configurations and without an assessment of the potential buggy
level of those configurations, the QA process (\eg debugging)
will require a great deal of effort from developers.
%
%
To deal with such large number of configurations, first, one will
eliminate the invalid configurations that violate the {\em constraints
  among the features} in the
system~\cite{Classen:2011:SMC:1985793.1985838,Classen:2010:MCL:1806799.1806850,Gruler:2008:MMC:1424547.1424557,kang1990feature,kastner2012virtual,Post:2008:CLV:1642931.1642971}. However,
the number of configurations after this step is still exponential.
To balance between bug detection rate and the number of configurations
to be examined, the {\em configuration selection} process is applied.
An example of selection algorithms is the \textit{k-way} combinatorial
approach~\cite{johansen-splc12,Marijan:2013:PPT:2491627.2491646,Oster:2010:AIP:1885639.1885658,Perrouin:2010:AST:1828417.1828490},
which considers the system under test as a blackbox and selects a
subset with at most $k$ features.~However, even with a small value of
$k$, \eg $k=6$, inspecting a very large number of selected
configurations without prioritizing the variants most likely having
defects is still inefficient.
%
Therefore, one would need a prioritization strategy to rank the
configurations to be examined.

The current state-of-the-art configuration prioritization algorithm is
the similarity-based configuration prioritization
(SP)~\cite{Al-Hajjaji:2014:SPS:2648511.2648532}.  Unfortunately, SP is
still ineffective in detecting feature-interaction bugs.
Let us illustrate this via our example. Table~\ref{kway} shows the
partial set of configurations chosen by 4-wise sampling algorithm and
prioritized by SP~\cite{Al-Hajjaji:2014:SPS:2648511.2648532}.
%
%
%
The variant, where \texttt{TWL4030\_CORE}, \texttt{IRQ\_DOMAIN,
  OF\_IRQ}, and \texttt{OF\_DEVICE} are enabled, and \texttt{SPARC} is
disabled, with the maximum number of features is selected to be
examined first by the SP algorithm. For the next configuration, the
configuration that has the minimum number of similar features compared
to the previously selected configuration is picked (\ie the one in
which \texttt{TWL4030\_CORE}, \texttt{IRQ\_DOMAIN}, \texttt{OF\_IRQ},
and \texttt{OF\_DEVICE} are disabled, and \texttt{SPARC} is enabled).
%
Although this second configuration is most dissimilar to the first
one, it does not contain the features whose interactions cause
violations, and there is no bug revealed by the second configuration.
As a result, by SP's strategy, the result is not an efficient order
for inspection because the aforementioned compile-time and run-time
errors are not detected until the 4$^{th}$ and 6$^{th}$ configurations
are inspected respectively. The configuration with both interaction
bugs would only be discovered via the 7$^{th}$ configuration. In our
experiment (will be presented in Section~\ref{sec:evaluation}), 36.2\%
of the feature-interaction bugs in the public benchmark, the
Variability Bugs Database (VBDb)~\cite{Abal:2014:VBL:2642937.2642990},
cannot be revealed until at least 10 configurations are inspected in
the resulting list ranked by the $SP$ approach.



\begin{table}[t]
\small
\centering
\caption{The configurations ordered by SP algorithm~\cite{Al-Hajjaji:2014:SPS:2648511.2648532} for Figure~\ref{example1}}
\label{kway}
\begin{tabular}{|l|l|l|l|l|l|}
\hline
\#          & \texttt{OF\_IRQ}        & \texttt{IRQ\_DOMAIN}   & \texttt{OF\_DEVICE}     & \begin{tabular}[c]{@{}l@{}}\texttt{TWL4030}\\ \texttt{\_CORE}\end{tabular}          & \texttt{SPARC} \\ \hline
1           & T  	      & T 		  & T  			& T  			& F                                            \\ \hline
2          	& F           & F          & F           	& F          	& T                                                    \\ \hline
3          & F           & T          & F          	& F           	& T                                                    \\ \hline
4           & \textbf{F}  & \textbf{T} & \textbf{T} & \textbf{T}   & \textbf{F}                                                     \\ \hline
5          & T           & T         & F           	& T           	& T                                                    \\ \hline
6           & \textbf{T}           & \textbf{T}         & \textbf{T}  & \textbf{T}       & \textbf{T}                         \\ \hline
7           & \textbf{F}	 & \textbf{T}& \textbf{T}	& \textbf{T}    & \textbf{T}                                                    \\ \hline
\end{tabular}
\end{table}


\subsection{Observations}


Let us consider the code in Figure~\ref{example1} with the two
following feature interactions that can cause the violations of
program semantics: 1) the declaration of the function
\texttt{of\_platform\_populate} in feature \texttt{L} (line 5) and its
use in \texttt{Z} (line 24), and 2) the assignment of \texttt{ops} in
feature \texttt{Y} (line 19) and its reference in \texttt{K} (line
22).  There are two potential bugs: 1) the use of the function
\texttt{of\_platform\_populate} without its declaration; and 2) the
reference to the variable \texttt{ops} without its initiation.
The configuration that enables \texttt{Z} and disables \texttt{L}
(\texttt{CONFIG\_OF\_DEVICE=T}, \texttt{CONFIG\_SPARC=T}) and enables
\texttt{K} and disables \texttt{Y} (\texttt{CONFIG\_TWL4030\_CORE=T},
\texttt{CONFIG\_OF\_IRQ=F}) should be inspected earlier to detect the
two bugs. Based on this observation, those interactions between
features should be comprehended to quickly discover these above
interaction bugs.
%
%
That motivates us to propose an approach that first analyzes the
source code to more precisely detect the potential interactions among
features, and then assesses the probabilities to be faulty of the
configurations to prioritize to inspect/test them in a more efficient
order.

\subsection{{\tool} Overview}


Detecting all interactions among features in a sound and complete
manner requires an analysis on all combinations of configuration
options. That is prohibitively expensive and impractical.
To deal with this problem, we statically analyze the source code to
{\em locally and heuristically identify the interactions between
  features via the shared program entities and the operations on
  them}.
For example, \texttt{L} shares the function
\texttt{of\_platform\_populate} with \texttt{Z} (which is declared on
line 5 and used on line 24) and \texttt{K} interacts with \texttt{Y}
via the variable \texttt{ops} (which is assigned on line 19 and
referred to on line 22).
Importantly, the operations such as declaration, assignment, or
references on the shared entities could become invalid when certain
features (via configuration options) are enabled or disabled.
%
%
As a consequence, that could lead to a violation.
For instance, a violation occurs when \texttt{CONFIG\_TWL4030\_CORE},
\texttt{CONFIG\_OF\_DEVICE}, and \texttt{CONFIG\_SPARC} are enabled
because the function \texttt{of\_platform\_populate} would be used
(\texttt{K} and \texttt{Z} are enabled) while its declaration is turned
off (\texttt{L} is disabled).
%
%
The other violation occurs in the case that
\texttt{CONFIG\_TWL4030\_\-CORE} is \texttt{true}, that enables
\texttt{K}, while \texttt{Y} is disabled as \texttt{CONFIG\_OF\_IRQ}
is disabled. This would induce the bug of \textit{dereferencing the
  \texttt{NULL} pointer} on variable \texttt{ops} (line 12).
With our strategy, the 7$^{th}$ variant in Table~\ref{kway} is more
suspicious than the 4$^{th}$, 6$^{th}$, and any other ones.
Generally, the suspiciousness of a variant is determined by the number
of violations that it might induce. Finally, a configuration can be
ranked according to its suspiciousness score, thus, we could create a
prioritization order of variants that maximizes fault detection~rate.




\section{Formulation}
\label{sec:formulation}

Let us formulate the problem of feature-interaction-aware
configuration prioritization.

\subsection{Program Entities and Operations}

In a program, we are interested in the program entities and the
operations performed on them.

\begin{definition}{({\bf Program Entity}).}
A program entity is a program element that is uniquely identified with
a \textit{name} and a \textit{scope}. The \textit{scope} and the
\textit{name} of an entity are used together as the identity of the
entity.
\end{definition}

In our formulation, we are interested in two types of program
entities: {\em variable} and {\em function}.
An entity is represented in the form of \texttt{[scope.ent\_name]},
where \texttt{scope} and \texttt{ent\_name} are the scope and the name
of the program entity respectively. For example, the code in
Figure~\ref{example1} contains the variables
\texttt{GLOBAL.irq\_domain\_simple\_ops}, \texttt{twl\_probe.ops},
the function \texttt{GLOBAL.twl\_probe}, \etc

We define 4 types of operations on variables and functions.


\begin{definition}{({\bf Program Operation}).}
We define four types of operations on variables and functions:
\textit{declare}, \textit{assign}, \textit{use} and
\textit{destruct}. Let $OP$ be the set of program operations,
$OP=\{declare, assign,$ $use, destruct\}$. All of those four
operations are applicable to variables, while $declare$
and $use$ are only applicable on functions.
\end{definition}

For variables, the \textit{assign} operation is used to assign a
non-null value to a variable. A \texttt{NULL} assignment to a variable
is treated as a special case of an assignment. In
Figure~\ref{example1}, function
\texttt{GLOBAL.of\_plat\-form\_populate\_probe} is {\em declared} at
line 5, and {\em used} at line 24. \texttt{twl\_probe.ops} is {\em
  declared} (line 17), {\em assigned} a value (line 19), and then {\em
  used/referred} to (line 22).

\subsection{Configurations and Features}

A configurable system contains several segments of code that are
present in any variant that implements its basic functionality. Those
segment form \textit{the core} of the system.

In practice, a configurable system usually provides a large number of
{\bf configuration options} to configure several optional segments of
code to be present or absent, in addition to the core of the system.
Those optional segments of code are aimed to realize the
optional \textbf{features} of the system.
For example, in the Linux Kernel, the configuration options have the
prefix of \texttt{CONFIG\_}, and they can have different
values. Without loss of generality, we assume that the value of a
configuration option is either \texttt{true(T)} or \texttt{false(F)}
(We can consider the entire conditional expressions of non-boolean
options as boolean ones, \eg \texttt{CONFIG\_A>10} as
\texttt{CONFIG\_A>10=T/F}).

\begin{definition}{({\bf Configuration Option}).}
A configuration option ({\em option} for short) is an element that is used
to configure the source code of a configurable system, such that the
option's value determines the presence or absence of one or more
segments of~code.
\end{definition}

In a configurable system, the presence or absence of code segments is
dependent on the values of multiple options. In Figure~\ref{example1},
the lines 19 and 20 are presented only when both
\texttt{CONFIG\_TWL4030\_CORE} and \texttt{CONFIG\_OF\_IRQ} are
\texttt{T}. Thus, at line 19, \texttt{irq\_domain\_simple\_ops} is
potentially used to assign as a value to the variable \texttt{ops}
when both of those options are \texttt{T}.


\begin{table*}[t]
\centering
\caption{Different Kinds of Feature Interactions}
\label{feature-interaction-types}
\begin{tabular}{|l|l|l|}
\hline
& Kind of Interaction & Condition \\ \hline

1&\textit{declare-declare}& $\exists e \in \rho_1 \cap \rho_2$, $e$ is declared in both $f_1$ and $f_2$\\ \hline
2&\textit{declare-assign}& $\exists e \in \rho_1 \cap \rho_2$, $e$ is declared in $f_1$ and then assigned in $f_2$\\ \hline
3&\textit{declare-use} & $\exists e \in \rho_1 \cap \rho_2$, $e$ is declared in $f_1$ and used in $f_2$\\ \hline
4&\textit{declare-destruct} & $\exists e \in \rho_1 \cap \rho_2$, $e$ is declared in $f_1$, and destructed in $f_2$\\ \hline
5&\textit{assign-assign} & $\exists e \in \rho_1 \cap \rho_2$, $e$ is assigned in both $f_1$ and $f_2$\\ \hline
6&\textit{assign-use} & $\exists e \in \rho_1 \cap \rho_2$, $e$ is assigned in $f_1$ and used in $f_2$\\ \hline
7&\textit{assign-destruct} & $\exists e \in \rho_1 \cap \rho_2$, $e$ is assign in $f_1$ and destructed in $f_2$\\ \hline
8&\textit{use-destruct} & $\exists e \in \rho_1 \cap \rho_2$, $e$ is used in $f_1$ and destructed in $f_2$\\ \hline
9&\textit{destruct-destruct} & $\exists e \in \rho_1 \cap \rho_2$, the entity is destructed in both $f_1$ and $f_2$\\ \hline
\end{tabular}
\end{table*}

\begin{definition}{({\bf Selection Functions}).}
In a configurable system, we define selection functions as the
functions from $O\times V$ to $2^P$, where $O$ is the set of
configuration options, $V=\{\texttt{T, F}\}$, and $P$ is the set of program
entities used in the code of the configurable system. We define four selection functions:

\begin{itemize}[leftmargin=*]

\item $\alpha: O \times V \to 2^P$, $\alpha(o, v) = D$, where $o \in O, v \in \{\texttt{T, F}\}$, and $D$ is the set of entities potentially {\bf declared} if $o = v$.

\item $\beta: O \times V \to 2^P$, $\beta(o, v) = D$, where $o \in O, v \in \{\texttt{T, F}\}$, and $D$ is the set of entities potentially {\bf assigned} if $o = v$.

\item $\gamma: O \times V \to 2^P$, $\gamma(o, v) = D$, where $o \in O, v \in \{\texttt{T, F}\}$, and $D$ is the set of entities potentially {\bf used} if $o = v$.

\item $\delta: O \times V \to 2^P$, $\delta(o, v) = D$, where $o \in O, v \in \{\texttt{T, F}\}$, and $D$ is the set of entities potentially {\bf destructed} if $o = v$.
\end{itemize}

\end{definition}

For example, in Figure~\ref{example1}:

\begin{itemize}

\item $\alpha($\texttt{CONFIG\_SPARC, F}$)$=$\{$\texttt{GLOBAL.of\_platform\_populate,$\\ $of\_platform\_populate.node}$\}$

\item $\beta($\texttt{CONFIG\_OF\_IRQ, T}$)$=$\{$\texttt{twl\_probe.ops}$\}$

\item $\gamma($\texttt{CONFIG\_OF\_IRQ, T} $)$=$\{$\texttt{GLOBAL.irq\_domain\_simple\_$\\$ops}$\}$

\end{itemize}

\begin{definition}{({\bf Configuration}).}
Given a configurable system, a configuration is a specific
selection of configuration options, which defines a \textbf{variant}
of the system.
\end{definition}

Configuration options are used to control the features that are
represented by certain segments of code. For example, in
Figure~\ref{example1}, the feature represented by the segment of code
\texttt{X} (feature \texttt{X}) is enabled if the value of
the configuration option \texttt{CONFIG\_IRQ\_DOMAIN} is \texttt{true},
whereas feature \texttt{Y} is enabled if both \texttt{CONFIG\_OF\_IRQ} and \texttt{CONFIG\_\-TWL4030\_CORE} are \texttt{true}.

\begin{definition}{({\bf Feature}).}
In a configurable system, a feature~$f$ is implemented by applying
program operations on a set of program entities, whose
presence/absence is controlled by certain configuration options.
%
%
We denote it by $f \sim OP \times \rho$ where $OP$ is the set of
program operations and $\rho$ is the set of program entities.
\end{definition}

A special case of features is that $f$ is \textit{the core feature}
($F$), $A \cup B \cup \Gamma \cup \Delta=\rho$, where $A, B, \Gamma,
\Delta$ are the sets of program entities that are declared, assigned,
used and destructed in the core system. $F$ is not controlled
by any configuration option.






\subsection{Feature Interactions}

In a configurable system, a feature may influence or~modify (often
called \textit{interact} with) the functions offered~by other
features through shared program entities that are used to implement
the features. For example, features~\texttt{X}, \texttt{K} and \texttt{Z} 
interact with one another via the variables \texttt{GLO\-BAL.irq\_domain\_simple\_ops} and \texttt{twl\_probe.temp}.
The manners the features interacting with each other depend on how the
shared entities are operated. For example, feature \texttt{Y}
\textit{assigns} \texttt{\&irq\_domain\_simple\_ops} to \texttt{ops}
and feature \texttt{K} \textit{uses} that variable (line 22). If no
assignment was done in \texttt{Y}, dereferencing in \texttt{K} would
be invalid, causing a \texttt{NULL} pointer~exception.


{\em Multi-way feature-interaction.} We present only on the
interactions between pairs of features because the interactions
between more than two features can be modeled as the operations on the
shared variables between pairs of features. Let us provide a sketch
of the proof for this statement.
We assume that there exists an interaction among $m$ features ($m >
2$). For simplicity, we consider the case of $m=3$, and the
interaction among $f_1\sim OP \times \rho_1, f_2\sim OP \times \rho_2$
and $f_3\sim OP \times \rho_3$. There are two cases of this
interaction. First, there exists an entity that shared by all 3
features, $\rho_1 \cap \rho_2 \cap \rho_3 = \omega \neq
\emptyset$. Since $\rho_1 \cap \rho_2 \supset \omega$ and $\rho_2 \cap
\rho_3 \supset \omega$, identifying interactions between pairs
directly captures the interaction among 3 features. The second case is
that $\rho_1 \cap \rho_2 = \omega_1$, $\rho_2 \cap \rho_3 = \omega_2$
and $\omega_1 \cap \omega_2 = \emptyset$. Meanwhile, $f_3$ is
influenced by $f_1$ (because the roles of $f_1$ and $f_3$ features in
this case are equal). This leads to that there exist entities: $e_1
\in \omega_1, e_2 \in \omega_2$, such that $e_2 = p(e_1)$, where $p$
is a value propagation function. This means the value of $e_1$ is
propagated to $e_2$, and that influences $f_3$. Hence, the interaction
among 3 features is still captured by determining interactions between
pairs of features.

For instance, the interaction among features \texttt{X}, \texttt{K} and \texttt{Z} can be broken
down into the shared program entities between two pairs of features as
follows: (\texttt{X}, \texttt{K}) via the variable \texttt{GLOBAL.irq\_domain\_\-simple\_ops}, and
(\texttt{K}, \texttt{Z}) via the variable \texttt{twl\_probe.temp}.
%
%
%
Thus, our solution can still model the interactions with more than two
features via the operations on their shared program entities.
From now on, we refer to a feature interaction as an
interaction determined via the shared program entities between
a pair of features.


In {\tool}, we focus on the feature interaction through the shared
program entities. The feature interactions when the variables are
associated with the external data such as when they interfere with
each other's behaviors on files or databases are beyond the scope of
our static analysis-based solution. Similarly, we will not detect the
interactions through pointers or arrays in this work.  As a
consequence, if both features {\em use} (refer to) a program entity,
they will not change the program's state. Thus, there is no
interaction between two features if they only use shared functions and
variables.

%

With the above design focuses, in {\tool}, the interactions between
two features $f_1 \sim OP \times \rho_1$, and $f_2 \sim OP \times
\rho_2$ with $\rho_1 \cap \rho_2 \neq \emptyset$, can be categorized
into nine kinds of interactions that are displayed in
Table~\ref{feature-interaction-types} (the \textit{use-use} case is
eliminated as explained).



  
  







\subsection{Feature Interaction Detection}


In a configurable system, the features (except the core features of
the system) are controlled by certain configuration options. Thus, if
there exists an interaction among the features, the interaction will
be one of the following:

\begin{itemize}

\item \textit{declare-declare}, there exist two option $o_1, o_2$ and their selected values $v_1, v_2$, such that $\alpha(o_1, v_1) \cap \alpha(o_2, v_2) \neq \emptyset$

\item \textit{declare-assign}, there exist two option $o_1, o_2$ and their selected values $v_1, v_2$, such that $\alpha(o_1, v_1) \cap \beta(o_2, v_2) \neq \emptyset$

\item \textit{declare-use}, there exist two option $o_1, o_2$ and their selected values $v_1, v_2$, such that $\alpha(o_1, v_1) \cap \gamma(o_2, v_2) \neq \emptyset$
  
\item \textit{declare-destruct}, there exist two option $o_1, o_2$ and their selected values $v_1, v_2$, such that $\alpha(o_1, v_1) \cap \delta(o_2, v_2) \neq \emptyset$

\item \textit{assign-assign}, there exist two option $o_1, o_2$ and their selected values $v_1, v_2$, such that $\beta(o_1, v_1) \cap \beta(o_2, v_2) \neq \emptyset$

\item \textit{assign-use}, there exist two option $o_1, o_2$ and their selected values $v_1, v_2$, such that $\beta(o_1, v_1) \cap \gamma(o_2, v_2) \neq \emptyset$

\item \textit{assign-destruct}, there exist two option $o_1, o_2$ and their selected values $v_1, v_2$, such that $\beta(o_1, v_1) \cap \delta(o_2, v_2) \neq \emptyset$

\item \textit{use-destruct}, there exist two option $o_1, o_2$ and their selected values $v_1, v_2$, such that $\gamma(o_1, v_1) \cap \delta(o_2, v_2) \neq \emptyset$

\item \textit{destruct-destruct}, there exist two option $o_1, o_2$ and their selected values $v_1, v_2$, such that $\delta(o_1, v_1) \cap \delta(o_2, v_2) \neq \emptyset$

\end{itemize}



Based on the above rules, our feature-interaction detection algorithm
statically analyzes the source code and configuration options, and
then computes the sets $\alpha$, $\beta$, $\gamma$, and $\delta$ for
any two options $o_1$ and $o_2$. For example, we can detect a
\textit{declare-declare} interaction between 2 features if there
exists 2 options $o_1$ and $o_2$, such that $\alpha(o_1, v_1) \cap
\alpha(o_2, v_2) \neq \emptyset$, where $v_1, v_2$ are their selected
values.  Other detection rules are similarly derived. For example,
because $\beta($\texttt{CONFIG\_OF\_IRQ, T}$) \cap
\gamma($\texttt{CONFIG\_TWL4030\_CORE, T}$)=\{$\texttt{ops}$\}$, there is a potential
\textit{assign-use} interaction among features. Thus, in this case, the actual \textit{assign-use} interaction
among \texttt{Y} and \texttt{K} exists.


For the core feature, if $F$ and other features interact with
one another, depending on the kinds of the interaction, there exists a
selection $v$ of an option $o$, such that $\alpha(o, v), \beta(o, v),
\gamma(o, v),\delta(o, v)$ intersect with $A, B, \Gamma, \Delta$,
\ie intersecting with the entities in the~core. Interactions among
core features and others are similarly identified.

In this version of {\tool}, we formulate feature interaction
statically through the completed set of operations on the entities
that are shared between features. More sophisticated interactions
relevant to pointers and external data such as files or databases can
be detected by using different data structures in the same principle
and using other types of analysis.

\section{Configuration Prioritization}



\begin{table*}[t]
\centering
\caption{Different kinds of feature-interaction defects}
\label{bugtypes}
\begin{tabular}{|l|l|l|l|l|}
\hline
&Kind of interaction & Detection condition & Suspicious selection & Potential violation \\ \hline

1&\textit{declare-declare}&$\alpha(o_1, v_1) \cap \alpha(o_2, v_2) \neq \emptyset$ &$o_1=v_1, o_2=v_2$&Declaration duplication\\ \hline
2&\textit{declare-use}&$\alpha(o_1, v_1) \cap \gamma(o_2, v_2) \neq \emptyset$&$o_1=v'_1, o_2=v_2$&Use without declaration\\ \hline
3&\textit{declare-use}&$\alpha(o_1, v_1) \cap \gamma(o_2, v_2) \neq \emptyset$&$o_1=v_1, o_2=v'_2$&Unused variables/functions\\ \hline
4&\textit{declare-destruct}&$\alpha(o_1, v_1) \cap \delta(o_2, v_2) \neq \emptyset$ &$o_1=v'_1, o_2=v_2$&Destruction without declaration\\ \hline
5&\textit{declare-assign}&$\beta(o_1, v_1) \cap \beta(o_2, v_2) \neq \emptyset$&$o_1=v_1, o_2=v_2$&Assignment without declaration\\ \hline
6&\textit{assign-use}&$\beta(o_1, v_1) \cap \gamma(o_2, v_2) \neq \emptyset$&$o_1=v'_1, o_2=v_2$&Use without assignment\\ \hline
7&\textit{assign-destruct}&$\beta(o_1, v_1) \cap \delta(o_2, v_2) \neq \emptyset$&$o_1=v'_1, o_2=v_2$&Destruction without definition\\ \hline
8&\textit{assign-destruct}&$\beta(o_1, v_1) \cap \delta(o_2, v_2) \neq \emptyset$&$o_1=v_1, o_2=v'_2$&Memory leak\\ \hline
9&\textit{destruct-destruct}&$\delta(o_1, v_1) \cap \delta(o_2, v_2) \neq \emptyset$&$o_1=v_1, o_2=v_2$&Destruction duplication\\ \hline
10&\textit{destruct-use}&$\delta(o_1, v_1) \cap \gamma(o_2, v_2) \neq \emptyset$&$o_1=v_1, o_2=v_2$& Use after destruction\\ \hline
\end{tabular}
\end{table*}

\subsection{Overview}

In general, to prioritize a given set of configurations under test,
our algorithm assigns a suspiciousness score to each configuration. The
suspiciousness score is determined via the number of the potential
feature-interaction bugs in different kinds that the variant
corresponding to that configuration might potentially have.

Feature-interaction bugs can be induced by any kinds of
interaction. Table~\ref{bugtypes} shows 10 different kinds of
feature-interaction bugs that are potentially caused by the respective
kinds of interactions listed in Table~\ref{feature-interaction-types} of
Section~\ref{sec:formulation}. The interactions in
Table~\ref{bugtypes} are called \textit{sensitive interactions} with
their \textit{suspicious selection} of options.
A configuration containing a suspicious selection potentially has the
corresponding violation.
%
%
For example, at line 6, if $\beta(o_1, v_1) \cap \gamma(o_2, v_2) \neq
\emptyset$, there is an \textit{assignment-use} potential interaction
between $f_1$ and $f_2$. When $o_1=v'_1, o_2=v_2$, where
$v'_1 \neq v_1$, $f_1$ might be disabled while $f_2$ is enabled, which
poses a violation of \textit{use without assignment}. In
Figure~\ref{example1}, because $\alpha($\texttt{CONFIG\_SPARC, F}$)
\cap \gamma($\texttt{CONFIG\_OF\_DEVICE,T}$) = \{$\texttt{GLOBAL.of\_platform\_populate}$\}$,
the program might not be compiled if \texttt{CONFIG\_SPARC} = \texttt{T} and
\texttt{CONFIG\_OF} \texttt{\_DEVICE} = \texttt{T} (\textit{use without declaration}).

\begin{algorithm}[t]
\caption{{\tool}: Feature-Interaction aware Configuration Prioritization Algorithm}	
 \label{agl1}
\begin{algorithmic}[1]
\Procedure {DetectSuspiciousSelections}{$Code$}
\State $Options = ExtractOptions(Code)$
\ForAll {$o \in Options$}
\State $TSelc = CollectProgramEntities(o, T, Code)$
\State $FSelc = CollectProgramEntities(o, F, Code)$
\State $Selections.add(TSelc)$
\State $Selections.add(FSelc)$
\EndFor

\ForAll {$selc \in Selections$}
	\ForAll {$other \in Selections$}
			\If {$ExistInteraction(selc, other)$}
				\If {$IsSensitiveInteraction(selc, other)$}
					\State $ss=ExtractSuspSelection(selc, other)$
					\State $SuspiciousSelections.add(ss)$
				\EndIf
			\EndIf
	\EndFor
\EndFor
\EndProcedure
\Statex

\Procedure {Prioritize}{$Configurations$, $SuspSelections$}
\ForAll {$c \in Configurations$}
	\State $SScore = CaculateSuspScore(c, SuspSelections)$
	\State $SetScore(c, SScore)$
\EndFor
\State $OrderBySuspiciousnessScoreDesc(Configurations)$
\EndProcedure
\end{algorithmic}
\end{algorithm}

\begin{table*}[t]
\centering
\caption{Configuration options and the values of 4 selection functions $\alpha$, $\beta$, $\gamma$, and $\delta$ for the example in Figure~\ref{example1}}
\label{foursets}
\begin{tabular}{|L{2cm}|L{1cm}|L{4.5cm}|L{2.5cm}|L{4.5cm}|L{0.5cm}|}
\hline
Option	& Value & $\alpha$	& $\beta$	& $\gamma$ 	& $\delta$ \\ \hline

\texttt{OF\_IRQ}       & \texttt{T}  &		& \texttt{twl\_probe.ops} & \texttt{GLOBAL.irq\_domain\_simple\_ops}	&         \\ \hline

\texttt{IRQ\_DOMAIN}   & \texttt{T}  & \texttt{GLOBAL.irq\_domain\_simple\_ops, GLOBAL.irq\_domain\_add, irq\_domain\_add.irq, irq\_domain\_add.ops} & \texttt{irq\_domain\_add.irq}   & \texttt{irq\_domain\_add.ops}	&         \\ \hline

\texttt{OF\_DEVICE}    & \texttt{T}  &                &                                                                     & \texttt{GLOBAL.of\_platform\_populate}	&         \\ \hline

\texttt{SPARC}         & \texttt{F} & \texttt{GLOBAL.of\_platform\_populate, of\_platform\_populate.node, of\_platform\_populate.t}	 &		& &         \\ \hline

\texttt{TWL4030\_CORE} & \texttt{T}  & \texttt{GLOBAL.twl\_probe, twl\_probe.n, twl\_probe.status, twl\_probe.temp, twl\_probe.ops}  & \texttt{twl\_probe.node, twl\_probe.temp, twl\_probe.status, twl\_probe.ops} & \texttt{GLOBAL.irq\_domain\_simple\_ops, GLOBAL.of\_platform\_populate, GLOBAL.irq\_domain\_add, twl\_probe.node, twl\_probe.temp, twl\_probe.status, twl\_probe.ops} &         \\ \hline
\end{tabular}
\end{table*}

\subsection{Detailed Algorithm}

The listing~\ref{agl1} shows the pseudo-code of {\tool}, our
feature-interaction aware configuration prioritization
algorithm. Given a configurable system, we first extract the set of
options used in the system. Then, for each selection $v$ of each
option $o$, the sets $\alpha(o, v), \beta(o, v), \gamma(o,v)$, and
$\delta(o,v)$ are computed via the function $CollectProgramEntities$ (lines 4--5). 
%
%
%
After that, for each
pair of option selections, it detects the potential interactions among
the features and checks whether the interactions are sensitive as
described in Table~\ref{bugtypes}.
Sensitive interactions are used to specify suspicious selections. This
information is used to compute the suspiciousness score for each
configuration after configuration selection (line 16). This score is
the number of suspicious selections contained by a configuration, and
equal to the number of potential bugs that the corresponding variant
might have. Finally, the configurations are ranked descendingly by
their suspiciousness~scores.

\subsection{Static Analysis}

In this version of {\tool}, to compute $\alpha, \beta, \gamma$, and
$\delta$ for the value $v$ of an option $o$ in
$CollectProgramEntities$, {\tool} analyzes the code by using
\textit{TypeChef}, a variability-aware
parser~\cite{Kenner:2010:TTT:1868688.1868693}. For a given
configurable code, \textit{TypeChef} is used to analyze and generate
the corresponding variability-aware control-flow graph. In a
variability-aware control-flow graph, the nodes refer to statements
and the edges, which are annotated with the corresponding presence
conditions, refer to the possible successor statements (conditional
statements). For the example in Figure~\ref{example1}, the successor
of the statement at line 22 is the conditional statement at line 24 if
\texttt{CONFIG\_OF\_DEVICE} is on, otherwise the statement at line 26
is the direct successor of the statement at line 22. After that,
{\tool} analyzes every conditional statements in the generated
control-flow graph to identify the entities that are either declared,
defined, used, or destructed in the statement and compute $\alpha,
\beta, \gamma$, and $\delta$ for the options and its values in the
corresponding presence conditions. For the statement at line 24 in
Figure \ref{example1}, if the value of \texttt{CONFIG\_OF\_DEVICE} is
\texttt{T}, the variable \texttt{status} is defined by using
\texttt{of\_platform\_populate} and \texttt{n}. This leads to that the
variable \texttt{status} is in $\beta($\texttt{CONFIG\_OF\_DEVICE}$,
$\texttt{T}$)$, and $\gamma($\texttt{CONFIG\_OF\_DEVICE}$,
$\texttt{T}$)$ contains the function \texttt{of\_platform\_populate} and
the variable~\texttt{n}.


\subsection{Running Example}

Let us illustrate our
algorithm via the example shown in Figure~\ref{example1}. {\tool}
computes the sets of the selection functions for each option, and the
result is shown in Table~\ref{foursets}. Based on the description on
Table~\ref{bugtypes}, the \textit{suspicious} selections include:


\begin{compactitem}

\item \texttt{CONFIG\_OF\_IRQ=F}, \texttt{CONFIG\_TWL4030\_CORE = T}

\item \texttt{CONFIG\_SPARC=T}, \texttt{CONFIG\_TWL4030\_CORE = T}

\item \texttt{CONFIG\_SPARC=T}, \texttt{CONFIG\_OF\_DEVICE = T}

\item \texttt{CONFIG\_IRQ\_DOMAIN=F}, \texttt{CONFIG\_OF\_IRQ = T}

\item \texttt{CONFIG\_IRQ\_DOMAIN=F}, \texttt{CONFIG\_TWL4030\_CORE = T}

\end{compactitem}

\begin{table}[t]
\centering
\tabcolsep 3pt
\small
\caption{Top-3 configurations ranked by {\tool} for Figure~\ref{example1}}
\label{ranking}
\begin{tabular}{|l|l|l|l|l|l|l|l|}
\hline
Rank by & Rank          & \texttt{OF\_}        & \texttt{IRQ\_}   & \texttt{OF\_}x     & \texttt{SPARC}          & \texttt{TWL4030\_}			&Score						\\ 
{\tool} & by SP &         \texttt{IRQ}      &        \texttt{DOMAIN}         &   \texttt{DEVICE}     &             & \texttt{CORE}   &  \\ \hline
1 & 7			& F 		& T 			& T  			& T  	   & T                   &3                         \\ \hline
2 & 6           & T  		 & T 		 & T  		  & T  		   & T                   &2                         \\ \hline
3 & 4           & T           & F         & T           & T           & F                  &2                                  \\ \hline
\end{tabular}
\end{table}

Based on the suspicious selections, {\tool} assigns the suspiciousness
scores and ranks all the configurations accordingly.
Table~\ref{ranking} shows the ranked configurations for our
example with their corresponding scores.
The top-ranked configuration by {\tool} is the 7$^{th}$ configuration
in the order generated by the \textit{ACTS} tool~\cite{actstool}, a
combinatorial test generation tool (see Table~\ref{kway}).  The
configuration covers both interaction bugs. Thus, after
inspecting/testing the first configuration, those two bugs will be
detected. In other words, {\tool} effectively ranks higher the
potential buggy variant than the SP algorithm.



\section{Empirical Evaluation}
\label{sec:evaluation}

To evaluate our configuration prioritization approach, we sought to answer the following:

\begin{enumerate}[label={\bf RQ{\arabic*}}]

\item \textbf{[Performance against a benchmark].} How does {\tool}
  perform on Variability Bugs Database
  (VBDb)~\cite{Abal:2014:VBL:2642937.2642990}, a public dataset of
  bugs in configurable code?



\item \textbf{[Comparison].}  How does {\tool} improve over the
  baseline random prioritization and similarity-based
  prioritization~\cite{Al-Hajjaji:2014:SPS:2648511.2648532} approaches
  when we add each of them on top of advanced sampling configuration
  selection algorithms?


\item \textbf{[Performance in the wild].} How does it perform on
  not-yet discovered interaction bugs in configurable systems?

\item {\bf [Time Complexity]} What is {\tool}'s running time?

\end{enumerate}

To answer RQ1 and RQ2, we conducted an experiment~to evaluate {\tool}
in a controlled environment with the VBDb public benchmark of
configuration-related
bugs~\cite{Abal:2014:VBL:2642937.2642990}. Answering RQ2 helps
evaluate how much improvement {\tool} gains over the {\em random
  prioritization} and the state-of-the-art {\em similarity-based
  prioritization}~\cite{Al-Hajjaji:2014:SPS:2648511.2648532}, when
adding {\tool} on top of the advanced configuration selection
techniques~\cite{Medeiros:2016:CSA:2884781.2884793}.
%
We answer RQ3 to evaluate {\tool} in the real-world setting. While the
bug detection tools cannot directly work on configurable code, with
{\tool}, we run them on the list of suspicious configurations ranked
by {\tool}.



\subsection{Subject Systems}

To evaluate {\tool}, we used two datasets in two different
experiments. To answer RQ1 and RQ2, we used the Variability Bugs
Database (VBDb)~\cite{Abal:2014:VBL:2642937.2642990} as a
benchmark. This publicly available bug database has 98 manually
verified configuration-related bugs in different versions of
highly-configurable systems: the Linux kernel~\cite{linux-kernel},
BusyBox~\cite{busybox}, Marlin~\cite{marlin}, and Apache
HTTPD~\cite{apache-httpd}.
Because the VBDb contains configuration-related bugs other
than feature-interaction ones, we kept only 46
feature-interaction bugs in those systems.
Table~\ref{subjectstat} shows their information including
the minimum and maximum numbers of configuration options (\code{MinOpt},
\code{MaxOpt}), the minimum and maximum numbers of files (\code{MinFile},
\code{MaxFile}), and the number of feature-interaction bugs (\code{Bugs}).

For the second experiment of RQ3, we selected an open-source
configurable system with a long history: libpng~\cite{vim} v0.89 with
40KLOC in 19 files and 80 options, and xterm~\cite{xterm} v2.24 with
50KLOC in 50 files, and 501 configuration options.

\begin{table}[]
\centering
\tabcolsep 4pt
\caption{Subject Systems in Variability Bugs Database}
\label{subjectstat}
\begin{tabular}{|l|ll|ll|l|}
\hline
Systems      &  MinOpt & MaxOpt & MinFile & MaxFile & \#Bugs\\ \hline
Linux        &   3463 &  5504    &    18886     &  34012  &  43   \\ \hline
Busybox      &   349  &   1449   &     236    &  799   &  18  \\ \hline
HTTPD     	 &   602  &    791   &     264    &  426  & 23  \\ \hline
Marlin       &   243  &   715    &       38  &    135 & 14 \\ \hline
\end{tabular}
\end{table}


\subsection{Experimental Procedure}

For each known buggy version of a subject system, we chose to include
the maximum number of files of 100 and the maximum number of inclusion
level of 3 (due to the limitation of the TypeChef
tool~\cite{kastner-oopsla11} that we used for variability-aware
parsing). We first applied a configuration selection process. That is,
to produce the sampled sets of configurations for each buggy version,
we ran sampling algorithms to select~a subset of configurations. For
each buggy system version and a particular sampling algorithm, we 
ran {\tool} on the set of configurations selected by a sampling algorithm.
For comparison, we ran the random prioritization and
similarity-based prioritization
techniques~\cite{Al-Hajjaji:2014:SPS:2648511.2648532} on the same configurations.




To evaluate {\tool} on detecting not-yet reported
interaction bugs in VBDb, we first ran it on a subject system
to achieve the ranked list of the configurations.  We also collected
and analyzed the sensitive interactions and potential suspicious
selections reported by our tool to detect unknown bugs.
For the top-ranked configurations in the list with the reported
potential suspicious interactions, we used a compiler 
to detect bugs.




\subsection{Evaluation Metric}

For evaluation, we adopted the \textit{Average Percentage Faults
  Detected (APFD)} \cite{962562}, a widely-used metric in evaluating
test prioritization techniques. \textit{APFD} is originally applied
for evaluating the average percentage bugs detected by a test
suite. In this work, since we used {\tool} with a bug detection tool,
we used \textit{APFD} to measure prioritization effectiveness in term
of the rate of bug detection of a configuration set, which is defined
by the following formula:
\[
APFD=1-\frac{\sum_{i=1}^{m} CF_i}{n \times m}+\frac{1}{2 \times n}
\]

where $n$ and $m$ denote the number of configurations and the number
of bugs, respectively. $CF_i$ is the smallest number of configurations
in the list, which is needed to be inspected to detect the $i^{th}$
bug. The \textit{APFD} score is from $0$ to $1$. For the fixed numbers
of faults and configurations, the higher \textit{APFD}, the
higher fault-detection rate and the better ranking order.

\subsection{Effectiveness and Comparison (RQ1 and RQ2)}



\subsubsection{{\bf Comparative Results}}

\begin{table}[t]
\tabcolsep 3pt
\centering
\caption{Average APFD for {\tool} versus \textit{SP} and \textit{Random} prioritization (added on top of advanced sampling algorithms)}
\label{apfd_table}
\begin{tabular}{|l|l|l|l|l|l|l|}
\hline
 & \multicolumn{3}{c|}{\textbf{APFD}} & \multicolumn{3}{c|}{\textbf{AVG Rank}} \\ \cline{2-7} 
\multirow{-2}{*}{} & \textit{Random} & \textit{SP} & {\tool} & \textit{Random} & \textit{SP} & {\tool} \\ \hline
Pairwise & 0.68 & 0.75 & \cellcolor[HTML]{C0C0C0}0.93 & 5.12 & 4.11 & \cellcolor[HTML]{C0C0C0}1.55 \\ \hline
Three-wise & 0.83 & 0.89 & \cellcolor[HTML]{C0C0C0}0.96 & 7.80 & 4.79 & \cellcolor[HTML]{C0C0C0}2.39 \\ \hline
Four-wise & 0.88 & 0.94 & \cellcolor[HTML]{C0C0C0}0.97 & 11.57 & 6.26 & \cellcolor[HTML]{C0C0C0}3.77 \\ \hline
Five-wise & 0.89 & 0.93 & \cellcolor[HTML]{C0C0C0}0.97 & 11.03 & 6.74 & \cellcolor[HTML]{C0C0C0}3.49 \\ \hline
One-enabled & 0.64 & 0.69 & \cellcolor[HTML]{C0C0C0}0.91 & 36.87 & 30.55 & \cellcolor[HTML]{C0C0C0}13.19 \\ \hline
One-disabled & 0.60 & 0.56 & \cellcolor[HTML]{C0C0C0}0.88 & 37.34 & 38.21 & \cellcolor[HTML]{C0C0C0}14.76 \\ \hline
\begin{tabular}[c]{@{}l@{}}Most-enabled\\ -disabled\end{tabular} & 0.52 & 0.55 & \cellcolor[HTML]{C0C0C0}0.57 & 1.70 & 1.43 & \cellcolor[HTML]{C0C0C0}1.43 \\ \hline
\begin{tabular}[c]{@{}l@{}}Statement\\ -coverage\end{tabular} & 0.61 & 0.57 & \cellcolor[HTML]{C0C0C0}0.88 & 37.30 & 38.25 & \cellcolor[HTML]{C0C0C0}17.80 \\ \hline
\end{tabular}
\end{table}

Table~\ref{apfd_table} shows the comparative results in term of the
average APFD and average rank (AVG Rank) between {\tool} and the
state-of-the-art prioritization methods, when we ran all of them on
the results of the advanced sampling
techniques~\cite{Medeiros:2016:CSA:2884781.2884793}. As seen, {\bf
  {\tool} achieves 2--32\% higher APFD ({\bf 14.9\%} on average)
  compared to \textit{SP} and 5--28\% higher ({\bf 17.8\%} on average)
  compared to \textit{Random} approach}. {\tool} also achieves much
better ranking compared to \textit{SP} and \textit{Random}. For
example, using {\tool} with \textit{One-disabled}, which is
recommended by the authors of
VBDb~\cite{Abal:2014:VBL:2642937.2642990}, the interaction bugs are
revealed after no more than 15 configurations on average in the
resulting ranked list by {\tool} are inspected, instead of more than
37 configurations in the lists prioritized by \textit{SP} and
\textit{Random}. Especially, in {\bf {\bf 78.0\%} of the
  cases, {\tool} ranks the buggy configurations at the top-3 positions
  in the list, while the \textit{SP} and \textit{Random} approaches
  rank them at the top-3 positions for only 41.3\% and 26.1\% of the
  cases}.

We can also see that {\tool} outperforms the \textit{SP} and \textit{Random}
prioritization techniques consistently on the resulting configurations
selected by various advanced sampling algorithms. That is, if one
uses {\tool} to rank the configurations selected by advanced
algorithms, the inspection order by {\tool} is better than those of the
\textit{SP} and \textit{Random} prioritization. Note that in the case of \textit{Most-enabled-disabled}~\cite{Medeiros:2016:CSA:2884781.2884793}, for each buggy system, there are only two configurations selected by the sampling algorithm, and 23 out of 46 bugs cannot be revealed by the selected set of configurations. That makes all three prioritization approaches do not perform well in this case and achieve nearly equal average APFDs and ranks.
In brief, {\em {\tool} is able to rank the buggy configuration in a much
higher rank than SP and Random approaches}. In other words, {\em if we add 
{\tool} as the prioritization technique on top of the most advanced sampling 
algorithms, we would achieve a more effective solution than adding other 
prioritization approaches upon the selection algorithms}.



\subsubsection{{\bf Further Analysis}}

We further studied the cases in which {\tool} correctly ranks
the buggy configurations at the top positions.
For the cases with correct ranking (1--3), we found that in 77\% 
(30 out of 39) of these bugs, the features interact with one another via shared program
entities.
Thus, our rules in Section~\ref{sec:formulation} are applicable to
detect the majority of feature-interaction bugs in the public VBDb
benchmark.


We also found an interesting scenario of {\bf indirect
  feature-interactions} that {\tool} detected. In some of those 30
cases, {\tool} identifies sensitive interactions among features
indirectly via entities. For example, variable \texttt{x} is
initialized in the feature controlled by option \texttt{A} with
\texttt{A=T}. \texttt{x} is assigned to \texttt{y} in the feature
enabled if the option \texttt{B} is on. Then, \texttt{y} is referred
to in another place that controlled by option \texttt{C},
\texttt{C=T}. In this case, if \texttt{A=F}, \texttt{B=T}, and
\texttt{C=T}, a \textit{null pointer exception} might be induced.  In
this case, since of the propagation of variables' values, the
interaction between two features controlled by \texttt{A} and
\texttt{C} can be captured by {\tool} via the feature controlled
by~\texttt{B}. Thus, the buggy configurations are ranked on the
top. This also indicates {\tool}'s capability in detecting
configurations containing {\bf bugs relevant to more than two
  features}.

\subsubsection{{\bf Examples on Feature-Interaction Bugs}}
Let us present the configuration-related bugs involving {\bf
  high-degree feature interactions} and the cases that {\tool}
detected the feature-interaction bugs {\bf not-yet-discovered} in
the VBDb benchmark.

\vspace{0.03in}
\noindent {\bf \underline{A bug involving 6 configuration options}.}
Figure~\ref{apache1} shows a bug in Apache HTTPD at commit
2124ff4. The bug is in the file \texttt{mod\_cgid.c}. In this example, the bug
is observed when \texttt{RLIMIT\_CPU}, \texttt{RLMIT\_NPROC},
\texttt{RLIMIT\_DATA}, \texttt{RLIMIT\_VMEM}, and \texttt{RLIMIT\_AS}
are disabled, while \texttt{RLIMIT\_NPROC} is enabled.
%
With the selections of the combinations of those options, the
field \texttt{limits} of any variable of the
type \texttt{cgid\_red\_t} (e.g. \texttt{req}) used in any features is
not declared (line 3).
Meanwhile, the filed \texttt{limits} is used
in \texttt{req.limits} on line 12 when
\texttt{RLIMIT\_NPROC} is enabled.
By identifying the suspicious interactions between the features
controlled by the pairs of \texttt{RLIMIT\_NPROC} and each of these
5 other options via the field \texttt{req.limits}, {\tool} specifies
that the selection that \texttt{RLIMIT\_NPROC} = \texttt{T}, \texttt{RLIMIT\_CPU} = \texttt{F},
\texttt{RLMIT\_NPROC} = \texttt{F}, \texttt{RLIMIT\_DATA} = \texttt{F}, \texttt{RLIMIT\_VMEM} = \texttt{F},
and \texttt{RLIMIT\_AS} = \texttt{F} is more suspicious than all other
selections containing those six configuration options.



\begin{figure}[t]
\begin{lstlisting}[label=example2,captionpos=b]
    typedef struct {
    #if defined (RLIMIT_CPU) || defined (RLMIT_NPROC) || defined (RLIMIT_DATA) || defined(RLIMIT_VMEM   ) || defined(RLIMIT_AS)
        cgid_rlimit_t limits;
    #endif
    } cgid_req_t;
    static apr_status_t send_req(){
        cgid_req_t req = {0};
    #if defined(RLIMIT_DATA) || defined(RLIMIT_VMEM) || defined(RLIMIT_AS)
        req.limits.limit_mem_set = 1;
    #endif
    #ifdef RLIMIT_NPROC
        req.limits.limit_nproc = 0;
    #endif
    }
\end{lstlisting}
\caption{A 6-way Feature-Interaction Bug in Apache Httpd}
\label{apache1}
\end{figure}


\vspace{0.03in}
\noindent {\bf \underline{Not-yet discovered feature-interaction
    bugs in VBDb}} {\bf \underline{bench\-mark}}.  Interestingly, while
using VBDb, we were able to use {\tool}
detect the interaction bugs that were neither discovered and
reported in those systems nor in VBDb. In total,
{\bf we found 17 such feature-interaction bugs including
12 \textit{using-without-declaration} bugs, 2
\textit{memory-leak} bugs, 2 \textit{declaration duplication}
bugs, and 1 \textit{dead code} issue}.

Figure~\ref{busybox2} shows 2 not-yet-discovered bugs:
a \textit{memory leak} issue and an \textit{assignment
without~declaration} bug at commit fac312d78bf (which
also~has \textit{use without declaration} bug and \textit{destruction
without declaration} bug).
The \textit{assignment without declaration} bug occurs only if
\texttt{BB\_FEATURE\_LS\_SORT\-FILES} = \texttt{F} and
\texttt{CONFIG\_FEATURE\_LS\_RECURSIVE} = \texttt{T}. In this case,
\texttt{dndirs} and \texttt{dnd} are not declared since lines 3--4 are not included, but they are used at lines
11 and 13. Moreover, \texttt{dnd} is destructed on line 15. This bug
was fixed at commit ea224be6aa8 (in almost 6 years later). 3
years after that, a \textit{memory leak} issue was reported and
fixed at commit~ffd4774ad25: as \texttt{CONFIG\_FEATURE\_LS\_RECUR\-SIVE} is disabled, the memory
controlled by \texttt{subdnp} is initialized at line 9 and not
released. With {\tool}, it would have been fixed earlier.
%

\begin{figure}[t]
\begin{lstlisting}[captionpos=b]
    void showdirs(struct dnode **dn, int ndirs){
    #ifdef BB_FEATURE_LS_SORTFILES
	    int dndirs;
	    struct dnode **dnd;
    #endif
	    subdnp = list_dir(dn[i]->fullname);
    #ifdef CONFIG_FEATURE_LS_RECURSIVE
	    dnd = splitdnarray(subdnp, nfiles);
	    dndirs = countsubdirs(subdnp, nfiles);
    #ifdef CONFIG_FEATURE_LS_SORTFILES
	    shellsort(dnd, dndirs);
    #endif
	    showdirs(dnd, dndirs);
	    free(dnd);
	    free(subdnp);
    #endif
    }
\end{lstlisting}
\caption{Two Not-yet-discovered Bugs in Busybox}
\label{busybox2}
\end{figure}


\vspace{0.03in}
\noindent {\bf \underline{A run-time  feature-interaction Bug in Busybox}}
{\tool} is also able to detect run-time errors caused by feature
interactions. Figure~\ref{busybox1} shows a simplified bug in Busybox
extracted from
http://vbdb.itu.dk/\#bug/busybox/061fd0a. In this case, a bug occurs when \texttt{CONFIG\_FEATURE\_HDPARM\_\-HDIO\_UNREGISTER\_HWIF} = 
\texttt{T} if \texttt{c}=\texttt{`U'} and \texttt{p} = \texttt{NULL}. The execution goes
to \texttt{expected\_hwif\_\-error}. However, this label is visible
only when \texttt{CONFIG\_FEATURE\_HD\-PARM\_HDIO\_\-SCAN\_HWIF} =
\texttt{T}. Otherwise, we would have a run-time~error.

\begin{figure}[t]
\begin{lstlisting}[captionpos=b]
    int main(int argc, char** argv){
      int r = rand() % 2;
      char *p;
      char c;
      scanf("%c", &c);
      switch (c){
        case 'W':
          if (r)
            p = *argv++, --argc;
          break;
     #ifdef CONFIG_FEATURE_HDPARM_HDIO_UNREGISTER_HWIF
        case 'U':
          if(!p)
             goto expected_hwif_error; //ERROR
          break;
     #endif /*CONFIG_FEATURE_HDPARM_HDIO_UNREGISTER_HWIF*/
     #ifdef CONFIG_FEATURE_HDPARM_HDIO_SCAN_HWIF
        case 'R':
          if(!p)
            goto expected_hwif_error;
     expected_hwif_error:
       printf("expected hwif value");

     #endif /* CONFIG_FEATURE_HDPARM_HDIO_SCAN_HWIF */
     }
     return 0;
   }
\end{lstlisting}
\caption{A Run-time Feature-Interaction Bug in Busybox}
\label{busybox1}
\end{figure}

\subsection{Effectiveness in Detecting Bugs in the Wild (RQ3)}

To evaluate the effectiveness of {\tool} on the real-world,
open-source projects, we ran it on the configurable systems
\textit{libpng} v0.89 and \textit{xterm} v2.24
 to detect interaction bugs. Interestingly, with {\tool}, we were able
 to detect {\bf 4 interaction bugs that have not been
   reported/discovered before}. They have the same nature of {\em
   using variables/functions without declarations}. Let us discuss two
 of them in details. The other one can be found on our
 website~\cite{website}.


In Figure \ref{libpng_bug}, the code contains 2 bugs. The first one is
observed when the option \texttt{PNG\_READ\_INTER\-LACING\-\_SUPPORTED} or
\texttt{PNG\_WRITE\_INTERLACING\_SUPPORTED} is enabled (line~4) and
\texttt{PNG\_INTERNAL} is disabled (line~1). In this case,
the function \texttt{png\_set\-\_interlace\_\-handling} is declared (line 5), and
\texttt{PNG\_INTERLACE} (line 6) is used inside this
function. Meanwhile, the constant \texttt{PNG\_INTERLACE} (line 2) is declared only
if \texttt{PNG\_INTERNAL} is enabled. Thus, if \texttt{PNG\_INTERNAL}
is disabled, and either \texttt{PNG\_READ\_INTERLACING\_SUPPORTED} or
\texttt{PNG\_WRITE\_INTERLACING\_SUPPORTED} is enabled, we will
have a compiling error at line 6. The second bug occurs when both 
\texttt{PNG\_READ\_INTERLACING\_\-SUPPORTED} and
\texttt{PNG\_WRITE\_INTERLACING\_SUPPORTED} are \texttt{F}. In this
case, \texttt{png\_\-read\_image} use an undeclared function
(line~10).

\begin{figure}[t]
\begin{lstlisting}[captionpos=b]
    #if defined(PNG_INTERNAL)
    #define PNG_INTERLACE          0x0002
    #endif /* PNG_INTERNAL */
    #if defined(PNG_READ_INTERLACING_SUPPORTED) || defined(PNG_WRITE_INTERLACING_SUPPORTED)
       int png_set_interlace_handling(png_structp png_ptr){
          png_ptr->transformations |= PNG_INTERLACE;
    }
    #endif
    void png_read_image(png_structp png_ptr){
       int pass = png_set_interlace_handling(png_ptr);
    }
\end{lstlisting}
\caption{Two Not-yet-discovered Bugs in libpng}
\label{libpng_bug}
\end{figure}

\subsection{Time Complexity (RQ4)}
We run our experiments on a computer with Intel Core i5 2.7GHz
processor, 8GB RAM. The running time to analyze the most complex
case that contains 43KLOC and 194 configuration options and rank 156
configurations is 211,020ms.

\subsection{Limitations and Potential Solutions}

For the cases that {\tool} did not rank well the buggy configurations,
we found that the majority of them are not in the kinds of
interaction-related defects listed in Section~\ref{sec:formulation}.
For example, a variable \texttt{x} is assigned a value \texttt{v} if
option~\texttt{A} is enabled, otherwise \texttt{x=v'}. Then,
\texttt{x} is referred to in a~feature controlled by option
\texttt{B}. In this case, {\tool} detects the interactions between
those features. However, as a static technique, {\tool} could not
conclude which option selections are more suspicious.  To overcome
such limitation, one could use a dynamic analysis approach for
configurable code~\cite{icse14-varex}.

%
Figure~\ref{apache2} shows a simplified bug in HTTPD
(commit 9327311d30f) that {\tool} did not rank well the buggy~configurations. In Figure~\ref{apache2}, a \textit{use without assignment} is
exposed when \texttt{APU\_HAS\_LDAP} and
\texttt{APU\_HAS\_SHARED\_MEMORY} are on. {\tool} did not work
since there is no feature where
\texttt{rmm\_lock} is assigned. Consequently, no \textit{assign-use}
interaction exists.

\subsection{Extension to {\tool}}

Generally, to detect more kinds of bug such as in the above example, one
can extend our set of conditions with the corresponding violations in
Table~\ref{bugtypes}.
%
One can define a new condition to detect this bug as follows: i)
$\alpha($\texttt{APU\_HAS\_LDAP}$, $\texttt{T} $) \cap
\gamma($\texttt{APU\_HAS\_SHARED\_MEMORY}$, $\texttt{T}$) \neq
\emptyset = \{$\texttt{util\_ldap\_cache\_init.\-rmm\_lock}$\}$ and
ii) there is no definition of \texttt{rmm\_lock} in its scope, which
is the function \texttt{util\_ldap\_cache\_init}.

Interestingly, note that for this buggy system, {\tool} ranked the
configuration to reveal another flaw of \textit{unused variable}
(\texttt{rmm\_lock}) when \texttt{APU\_HAS\_LDAP=T} and
\texttt{APU\_HAS\_SHARED\_MEMORY=F}.

In 14 cases out of 368 cases, the interactions that cause the
interaction bugs are really detected, but the configurations that
reveal the bugs are still ranked lower than others. The reason for
these cases is that other configurations containing more suspicious
selections that actually do not cause the bugs. To faster detect the
bug in these situations, one can apply the \textit{Additional
  Priortization} strategy~\cite{test_prioritization} to rank the set
of configurations according to their numbers of potential bugs in an
incremental manner. By this strategy, the next configuration to
be~selected is the one containing the largest number of potential bugs
that have not been contained by the previous selected configurations
in the previous steps. Moreover, for the interaction bugs relevant to
pointers and external data files, one can define
new rules to add to our framework.

\begin{figure}[t]
\begin{lstlisting}[label=example_bad,captionpos=b]
    void apr_rmm_init(char* rmm_lock){
      printf("%s\n", rmm_lock);
    }
    #ifdef APU_HAS_LDAP
    void util_ldap_cache_init(){
       char* rmm_lock;
    #ifdef APR_HAS_SHARED_MEMORY
       apr_rmm_init(rmm_lock); // ERROR: rmm_lock uninitialized
    #endif
    }
    #endif
\end{lstlisting}
\caption{{\tool} did not rank well buggy configurations}
\label{apache2}
\end{figure}



\section{Related Work}
\label{related}

{\tool} is most closely related to the
work by Al-Hajjaji {\em et
  al.}~\cite{Al-Hajjaji:2014:SPS:2648511.2648532} on {\bf similarity-based
Prioritization} (SP). The key idea of SP approach dissimilar test sets
are likely to detect more defects than similar
ones~\cite{Al-Hajjaji:2014:SPS:2648511.2648532}. In SP, the
configuration with the maximum number of features is selected to be
the first one under test and the next configuration must have the
minimum number of similar features as the previously selected
configuration. In comparison, SP does not analyze the nature of
feature interactions, while {\tool} does. This avoids the problem in
SP that the different features to be considered next might not be the
ones that potentially causes violations.

{\tool} is also related to the work on {\bf configuration selection}
approaches to reduce the number of configurations~to be
tested~\cite{Medeiros:2016:CSA:2884781.2884793}. They focus on the
step before configuration prioritization, therefore the resulting set
of configurations is not ranked as in {\tool}.
The $t$-wise (i.e., $k$-way) sampling algorithm covers all
combinations of $t$
options~\cite{johansen-splc12,Lei:2008:IET:1405567.1405569,Nie:2011:SCT:1883612.1883618,Perrouin:2010:AST:1828417.1828490},
while pair-wise checks all pairs of configuration
options~\cite{Marijan:2013:PPT:2491627.2491646,Oster:2010:AIP:1885639.1885658}.
Recent study by Medeiros {\em et
  al.}~\cite{Medeiros:2016:CSA:2884781.2884793} showed that realistic
constraints among options, global analysis, header
files, and build-system information influence the performance of most
sampling algorithms substantially; and several algorithms are no
longer feasible in practice.
Importantly, they lack configuration prioritization, thus, developers
need to spend efforts to perform QA on all the variants.

{\tool} is also related to {\bf Variability-aware (VA)
  analysis}~\cite{Liebig:2013:SAV:2491411.2491437}. VA analysis is a
variation of a traditional analysis that considers the variability in
the configurable code.  The variability-aware analysis techniques have
been proposed for type checking~\cite{CEW:TOPLAS13,KATS:TOSEM12,Liebig:2013:SAV:2491411.2491437,TBKC:GPCE07}, model
checking~\cite{CHSLR:ICSE10,LTP:ASE09}, data-flow
analysis~\cite{spllift,BRTB:AOSD12,Liebig:2013:SAV:2491411.2491437},
and other analyses~\cite{CP:GPCE06} on multiple compile-time
configurations of a system at a time.
The main drawback of this approach is that it cannot reuse existing
static analysis tools, and each type of analysis must be rewritten in
a variability-aware fashion. For example, to detect \texttt{NULL}
exception, one must rewrite such an analysis to consider all different
configurations in a configurable code. In our experiment, we connect
{\tool} with an existing bug detection tool to work on configurable
code.  Variability-aware
execution~\cite{icse14-varex,Meinicke:2016:ECC:2970276.2970322}
explores multiple paths of execution at the same time to
detect feature-interaction bugs. However, it suffers scalability
issue.

Several approaches were proposed to detect {\bf feature
interactions}~\cite{Apel:2013:EFI:2528265.2528267,Garvin:2011:FIF:2120102.2120918}.
Verification~\cite{Apel:2011:DFI:2190078.2190192} is also used to detect feature-interaction
bugs.
Other prioritization approaches aim for {\em statement
  coverage}~\cite{tartler2014static,Tartler:2012:CCA:2094091.2094095}
via static checkers.
The issue is that computing an optimal solution for the coverage
problem is NP-hard, and including each block of optional code at least
once does not guarantee that all possible combinations of individual
blocks of optional code are
considered~\cite{Medeiros:2016:CSA:2884781.2884793}. To avoid finding
optimal coverage solution, the {\em
  most-enabled-disabled}~\cite{tartler2014static} algorithm checks two
samples independently of the number of configuration options. When
there are no constraints among configuration options, it enables all
options and then it disables all configuration
options. One-(enabled/disabled)
algorithm~\cite{Abal:2014:VBL:2642937.2642990} enables/disables one
configuration option at a time.
Despite different levels of heuristics, they do not analyze the
entities in source code.


Several pproaches are aimed for {\bf testing for configurable
systems}~\cite{Cabral:2010:ITT:1885639.1885662,Cohen:2007:ITH:1273463.1273482,Greiler:2012:TCS:2337223.2337253,Machado:2014:STS:2658281.2658306}. In
product-line testing~\cite{SPLEbook} and framework
testing~\cite{CEW:OSU12} it is a common strategy to unit test
components or plug-ins in isolation, while integration tests are often
neglected or performed only for specific configurations. Greiler {\em
  et al.} suggest shipping test cases with plug-ins and running them
in client systems~\cite{Greiler:2012:TCS:2337223.2337253}. In essence,
this postpones tests of configurations until the
configuration is actually used.

Other approaches have been proposed for static analysis of
product
lines~\cite{spllift,BRTB:AOSD12,CEW:TOPLAS13,CHSLR:ICSE10,CP:GPCE06,KATS:TOSEM12,TBKC:GPCE07,TAKKSS:MD12}. Researchers explore to represent and reason about partial but finite
configuration spaces compactly with BDDs or SAT solvers (as used in
our variability contexts)~\cite{B:SPLC05,KATS:TOSEM12,MWC:SPLC09},
choices of structures~\cite{EW:TOSEM10} and complex
structures~\cite{EW:GTTSE11,Liebig:2013:SAV:2491411.2491437}.

\section{Conclusion}


We propose {\tool}, a novel formulation of
feature-interaction bugs via common program entities enabled/disabled
by the features. Leveraging from that, we develop efficient
feature-interaction-aware configuration prioritization technique for a
configurable system by ranking the configurations according to their
total number of potential bugs. We
evaluated {\tool} in two complementary settings: detecting
configura\-tion-related bugs in a benchmark and a real-world open-source
systems. {\tool} outperforms the other techniques in
which in 78\% of the cases, it ranks the buggy configurations at the
top 3 positions. Interestingly, it is able to detect 17
not-yet-discovered, high-degree, feature-interaction~bugs.




\section*{Acknowledgment}
This work was supported in part by the US National Science
Foundation (NSF) grants CCF-1723215, CCF-1723432, TWC-1723198,
CCF-1518897, and CNS-1513263.



%



\balance

\bibliographystyle{plain}
\bibliography{references,splc}

\begin{thebibliography}{10}

\bibitem{website}
{}.
\newblock https://doubledoubleblind.github.io/copro/.

\bibitem{Abal:2014:VBL:2642937.2642990}
Iago Abal, Claus Brabrand, and Andrzej Wasowski.
\newblock {42 Variability Bugs in the Linux Kernel: A Qualitative Analysis}.
\newblock In {\em Proceedings of the 29th ACM/IEEE International Conference on
  Automated Software Engineering}, ASE '14, pages 421--432, New York, NY, USA,
  2014. ACM.

\bibitem{Al-Hajjaji:2014:SPS:2648511.2648532}
Mustafa Al-Hajjaji, Thomas Th\"{u}m, Jens Meinicke, Malte Lochau, and Gunter
  Saake.
\newblock Similarity-based prioritization in software product-line testing.
\newblock In {\em Proceedings of the 18th International Software Product Line
  Conference - Volume 1}, SPLC '14, pages 197--206, New York, NY, USA, 2014.
  ACM.

\bibitem{apache-httpd}
{Apache Httpd}.
\newblock http://httpd.apache.org/.

\bibitem{apel8overview}
Sven Apel and Christian K{\"a}stner.
\newblock An overview of feature-oriented software development.
\newblock {\em JOURNAL OF OBJECT TECHNOLOGY}, 8(5).

\bibitem{Apel:2013:EFI:2528265.2528267}
Sven Apel, Sergiy Kolesnikov, Norbert Siegmund, Christian K\"{a}stner, and
  Brady Garvin.
\newblock Exploring feature interactions in the wild: The new
  feature-interaction challenge.
\newblock In {\em Proceedings of the 5th International Workshop on
  Feature-Oriented Software Development}, FOSD '13, pages 1--8, New York, NY,
  USA, 2013. ACM.

\bibitem{Apel:2011:DFI:2190078.2190192}
Sven Apel, Hendrik Speidel, Philipp Wendler, Alexander von Rhein, and Dirk
  Beyer.
\newblock {Detection of Feature Interactions Using Feature-aware Verification}.
\newblock In {\em Proceedings of the 26th IEEE/ACM International Conference on
  Automated Software Engineering}, ASE '11, pages 372--375, Washington, DC,
  USA, 2011. IEEE Computer Society.

\bibitem{B:SPLC05}
Don Batory.
\newblock Feature models, grammars, and propositional formulas.
\newblock In {\em Proc.\ Int'l Software Product Line Conference (SPLC)}, volume
  3714 of {\em Lecture Notes in Computer Science}, pages 7--20,
  Berlin/Heidelberg, 2005. Springer-Verlag.

\bibitem{6572787}
T.~Berger, S.~She, R.~Lotufo, A.~Wasowski, and K.~Czarnecki.
\newblock A study of variability models and languages in the systems software
  domain.
\newblock {\em IEEE Transactions on Software Engineering}, 39(12):1611--1640,
  Dec 2013.

\bibitem{Berger:2013:SVM:2430502.2430513}
Thorsten Berger, Ralf Rublack, Divya Nair, Joanne~M. Atlee, Martin Becker,
  Krzysztof Czarnecki, and Andrzej Wk{a}sowski.
\newblock A survey of variability modeling in industrial practice.
\newblock In {\em Proceedings of the Seventh International Workshop on
  Variability Modelling of Software-intensive Systems}, VaMoS '13, pages
  7:1--7:8, New York, NY, USA, 2013. ACM.

\bibitem{spllift}
Eric Bodden, T\'{a}rsis Tol\^{e}do, M\'{a}rcio Ribeiro, Claus Brabrand, Paulo
  Borba, and Mira Mezini.
\newblock Spllift: Statically analyzing software product lines in minutes
  instead of years.
\newblock In {\em Proc.\ Conf.\ Programming Language Design and Implementation
  (PLDI)}, pages 355--364, New York, 2013. ACM Press.

\bibitem{BRTB:AOSD12}
Claus Brabrand, M\'arcio Ribeiro, T\'{a}rsis Tol\^{e}do, and Paulo Borba.
\newblock Intraprocedural dataflow analysis for software product lines.
\newblock In {\em Proc.\ Int'l Conf.\ Aspect-Oriented Software Development
  (AOSD)}, pages 13--24, New York, 2012. ACM Press.

\bibitem{busybox}
{Busy Box}.
\newblock https://busybox.net/.

\bibitem{Cabral:2010:ITT:1885639.1885662}
Isis Cabral, Myra~B. Cohen, and Gregg Rothermel.
\newblock Improving the testing and testability of software product lines.
\newblock In {\em Proceedings of the 14th International Conference on Software
  Product Lines: Going Beyond}, SPLC'10, pages 241--255, Berlin, Heidelberg,
  2010. Springer-Verlag.

\bibitem{CEW:OSU12}
Sheng Chen, Martin Erwig, and Eric Walkingshaw.
\newblock Extending type inference to variational programs.
\newblock Technical report (draft), School of EECS, Oregon State University,
  2012.

\bibitem{CEW:TOPLAS13}
Sheng Chen, Martin Erwig, and Eric Walkingshaw.
\newblock Extending type inference to variational programs.
\newblock {\em ACM Trans. Program. Lang. Syst. (TOPLAS)}, 2013.

\bibitem{Classen:2011:SMC:1985793.1985838}
Andreas Classen, Patrick Heymans, Pierre-Yves Schobbens, and Axel Legay.
\newblock Symbolic model checking of software product lines.
\newblock In {\em Proceedings of the 33rd International Conference on Software
  Engineering}, ICSE '11, pages 321--330, New York, NY, USA, 2011. ACM.

\bibitem{Classen:2010:MCL:1806799.1806850}
Andreas Classen, Patrick Heymans, Pierre-Yves Schobbens, Axel Legay, and
  Jean-Fran\c{c}ois Raskin.
\newblock Model checking lots of systems: Efficient verification of temporal
  properties in software product lines.
\newblock In {\em Proceedings of the 32nd ACM/IEEE International Conference on
  Software Engineering - Volume 1}, ICSE '10, pages 335--344, New York, NY,
  USA, 2010. ACM.

\bibitem{CHSLR:ICSE10}
Andreas Classen, Patrick Heymans, Pierre-Yves Schobbens, Axel Legay, and
  Jean-Francois Raskin.
\newblock Model checking lots of systems: Efficient verification of temporal
  properties in software product lines.
\newblock In {\em Proc.\ Int'l Conf.\ Software Engineering (ICSE)}, pages
  335--344, New York, 2010. ACM Press.

\bibitem{Cohen:2007:ITH:1273463.1273482}
Myra~B. Cohen, Matthew~B. Dwyer, and Jiangfan Shi.
\newblock Interaction testing of highly-configurable systems in the presence of
  constraints.
\newblock In {\em Proceedings of the 2007 International Symposium on Software
  Testing and Analysis}, ISSTA '07, pages 129--139, New York, NY, USA, 2007.
  ACM.

\bibitem{CP:GPCE06}
Krzysztof Czarnecki and Krzysztof Pietroszek.
\newblock Verifying feature-based model templates against well-formedness {OCL}
  constraints.
\newblock In {\em Proc.\ Int'l Conf.\ Generative Programming and Component
  Engineering (GPCE)}, pages 211--220, New York, 2006. ACM.

\bibitem{test_prioritization}
S.~Elbaum, A.~G. Malishevsky, and G.~Rothermel.
\newblock Test case prioritization: a family of empirical studies.
\newblock {\em IEEE Transactions on Software Engineering}, 28(2):159--182, Feb
  2002.

\bibitem{EW:TOSEM10}
Martin Erwig and Eric Walkingshaw.
\newblock The choice calculus: A representation for software variation.
\newblock {\em ACM Trans. Softw. Eng. Methodol. (TOSEM)}, 21(1):6:1--6:27,
  2011.

\bibitem{EW:GTTSE11}
Martin Erwig and Eric Walkingshaw.
\newblock Variation programming with the choice calculus.
\newblock In {\em Generative and Transformational Techniques in Software
  Engineering IV}, pages 55--100. Springer Berlin Heidelberg, 2013.

\bibitem{Garvin:2011:FIF:2120102.2120918}
Brady~J. Garvin and Myra~B. Cohen.
\newblock Feature interaction faults revisited: An exploratory study.
\newblock In {\em Proceedings of the 2011 IEEE 22nd International Symposium on
  Software Reliability Engineering}, ISSRE '11, pages 90--99, Washington, DC,
  USA, 2011. IEEE Computer Society.

\bibitem{Greiler:2012:TCS:2337223.2337253}
Michaela Greiler, Arie~van Deursen, and Margaret-Anne Storey.
\newblock Test confessions: A study of testing practices for plug-in systems.
\newblock In {\em Proceedings of the 34th International Conference on Software
  Engineering}, ICSE '12, pages 244--254, Piscataway, NJ, USA, 2012. IEEE
  Press.

\bibitem{Gruler:2008:MMC:1424547.1424557}
Alexander Gruler, Martin Leucker, and Kathrin Scheidemann.
\newblock Modeling and model checking software product lines.
\newblock In {\em Proceedings of the 10th IFIP WG 6.1 International Conference
  on Formal Methods for Open Object-Based Distributed Systems}, FMOODS '08,
  pages 113--131, Berlin, Heidelberg, 2008. Springer-Verlag.

\bibitem{johansen-splc12}
Martin~Fagereng Johansen, Oystein Haugen, and Franck Fleurey.
\newblock An algorithm for generating t-wise covering arrays from large feature
  models.
\newblock In {\em Proceedings of the 16th International Software Product Line
  Conference - Volume 1}, SPLC '12, pages 46--55, New York, NY, USA, 2012. ACM.

\bibitem{kang1990feature}
Kyo~C Kang, Sholom~G Cohen, James~A Hess, William~E Novak, and A~Spencer
  Peterson.
\newblock Feature-oriented domain analysis (foda) feasibility study.
\newblock Technical report, Carnegie-Mellon Univ Pittsburgh Pa Software
  Engineering Inst, 1990.

\bibitem{kastner2012virtual}
Christian K{\"a}stner.
\newblock Virtual separation of concerns: toward preprocessors 2.0.
\newblock {\em it-Information Technology Methoden und innovative Anwendungen
  der Informatik und Informationstechnik}, 54(1):42--46, 2012.

\bibitem{KATS:TOSEM12}
Christian K{\"a}stner, Sven Apel, Thomas Th{\"u}m, and Gunter Saake.
\newblock Type checking annotation-based product lines.
\newblock {\em ACM Trans. Softw. Eng. Methodol. (TOSEM)}, 21(3):14:1--14:39,
  2012.

\bibitem{kastner-oopsla11}
Christian K\"{a}stner, Paolo~G. Giarrusso, Tillmann Rendel, Sebastian Erdweg,
  Klaus Ostermann, and Thorsten Berger.
\newblock Variability-aware parsing in the presence of lexical macros and
  conditional compilation.
\newblock In {\em Proceedings of the 2011 ACM International Conference on
  Object Oriented Programming Systems Languages and Applications}, OOPSLA '11,
  pages 805--824, New York, NY, USA, 2011. ACM.

\bibitem{Kenner:2010:TTT:1868688.1868693}
Andy Kenner, Christian K\"{a}stner, Steffen Haase, and Thomas Leich.
\newblock {TypeChef: Toward Type Checking \#Ifdef Variability in C}.
\newblock In {\em Proceedings of the 2nd International Workshop on
  Feature-Oriented Software Development}, FOSD '10, pages 25--32, New York, NY,
  USA, 2010. ACM.

\bibitem{LTP:ASE09}
Kim Lauenroth, Klaus Pohl, and Simon Toehning.
\newblock Model checking of domain artifacts in product line engineering.
\newblock In {\em Proc.\ Int'l Conf.\ Automated Software Engineering (ASE)},
  pages 269--280, Los Alamitos, CA, 2009. IEEE Computer Society.

\bibitem{Lei:2008:IET:1405567.1405569}
Yu~Lei, Raghu Kacker, D.~Richard Kuhn, Vadim Okun, and James Lawrence.
\newblock Ipog-ipog-d: Efficient test generation for multi-way combinatorial
  testing.
\newblock {\em Softw. Test. Verif. Reliab.}, 18(3):125--148, September 2008.

\bibitem{vim}
{libpng}.
\newblock http://www.libpng.org/.

\bibitem{Liebig:2013:SAV:2491411.2491437}
J\"{o}rg Liebig, Alexander von Rhein, Christian K\"{a}stner, Sven Apel, Jens
  D\"{o}rre, and Christian Lengauer.
\newblock Scalable analysis of variable software.
\newblock In {\em Proceedings of the 2013 9th Joint Meeting on Foundations of
  Software Engineering}, ESEC/FSE 2013, pages 81--91, New York, NY, USA, 2013.
  ACM.

\bibitem{linux-kernel}
{Linux Kernel}.
\newblock https://www.kernel.org/.

\bibitem{Machado:2014:STS:2658281.2658306}
Ivan Do~Carmo Machado, John~D. Mcgregor, Yguarat\~{a}~Cerqueira Cavalcanti, and
  Eduardo~Santana De~Almeida.
\newblock On strategies for testing software product lines: A systematic
  literature review.
\newblock {\em Inf. Softw. Technol.}, 56(10):1183--1199, October 2014.

\bibitem{Marijan:2013:PPT:2491627.2491646}
Dusica Marijan, Arnaud Gotlieb, Sagar Sen, and Aymeric Hervieu.
\newblock Practical pairwise testing for software product lines.
\newblock In {\em Proceedings of the 17th International Software Product Line
  Conference}, SPLC '13, pages 227--235, New York, NY, USA, 2013. ACM.

\bibitem{marlin}
{Marlin}.
\newblock http://marlinfw.org/.

\bibitem{Medeiros:2016:CSA:2884781.2884793}
Fl\'{a}vio Medeiros, Christian K\"{a}stner, M\'{a}rcio Ribeiro, Rohit Gheyi,
  and Sven Apel.
\newblock A comparison of 10 sampling algorithms for configurable systems.
\newblock In {\em Proceedings of the 38th International Conference on Software
  Engineering}, ICSE '16, pages 643--654, New York, NY, USA, 2016. ACM.

\bibitem{Meinicke:2016:ECC:2970276.2970322}
Jens Meinicke, Chu-Pan Wong, Christian K\"{a}stner, Thomas Th\"{u}m, and Gunter
  Saake.
\newblock On essential configuration complexity: Measuring interactions in
  highly-configurable systems.
\newblock In {\em Proceedings of the 31st IEEE/ACM International Conference on
  Automated Software Engineering}, ASE 2016, pages 483--494, New York, NY, USA,
  2016. ACM.

\bibitem{MWC:SPLC09}
Marc\'{\i}lio Mendon\c{c}a, Andrzej Wk{a}sowski, and Krzysztof Czarnecki.
\newblock {SAT}-based analysis of feature models is easy.
\newblock In {\em Proc.\ Int'l Software Product Line Conference (SPLC)}, pages
  231--240, New York, 2009. ACM Press.

\bibitem{icse14-varex}
Hung~Viet Nguyen, Christian K\"{a}stner, and Tien~N. Nguyen.
\newblock Exploring variability-aware execution for testing plugin-based web
  applications.
\newblock In {\em Proceedings of the 36th International Conference on Software
  Engineering}, ICSE 2014, pages 907--918, New York, NY, USA, 2014. ACM.

\bibitem{Nie:2011:SCT:1883612.1883618}
Changhai Nie and Hareton Leung.
\newblock A survey of combinatorial testing.
\newblock {\em ACM Comput. Surv.}, 43(2):11:1--11:29, February 2011.

\bibitem{Oster:2010:AIP:1885639.1885658}
Sebastian Oster, Florian Markert, and Philipp Ritter.
\newblock Automated incremental pairwise testing of software product lines.
\newblock In {\em Proceedings of the 14th International Conference on Software
  Product Lines: Going Beyond}, SPLC'10, pages 196--210, Berlin, Heidelberg,
  2010. Springer-Verlag.

\bibitem{Perrouin:2010:AST:1828417.1828490}
Gilles Perrouin, Sagar Sen, Jacques Klein, Benoit Baudry, and Yves~le Traon.
\newblock Automated and scalable t-wise test case generation strategies for
  software product lines.
\newblock In {\em Proceedings of the 2010 Third International Conference on
  Software Testing, Verification and Validation}, ICST '10, pages 459--468,
  Washington, DC, USA, 2010. IEEE Computer Society.

\bibitem{SPLEbook}
Klaus Pohl, G\"{u}nter B\"{o}ckle, and Frank~J. van~der Linden.
\newblock {\em Software Product Line Engineering: Foundations, Principles and
  Techniques}.
\newblock Springer-Verlag, Berlin/Heidelberg, 2005.

\bibitem{Post:2008:CLV:1642931.1642971}
H.~Post and C.~Sinz.
\newblock Configuration lifting: Verification meets software configuration.
\newblock In {\em Proceedings of the 2008 23rd IEEE/ACM International
  Conference on Automated Software Engineering}, ASE '08, pages 347--350,
  Washington, DC, USA, 2008. IEEE Computer Society.

\bibitem{962562}
G.~Rothermel, R.~H. Untch, Chengyun Chu, and M.~J. Harrold.
\newblock Prioritizing test cases for regression testing.
\newblock {\em IEEE Transactions on Software Engineering}, 27(10):929--948, Oct
  2001.

\bibitem{tartler2014static}
Reinhard Tartler, Christian Dietrich, Julio Sincero, Wolfgang
  Schr{\"o}der-Preikschat, and Daniel Lohmann.
\newblock Static analysis of variability in system software: The 90,000\#
  ifdefs issue.

\bibitem{Tartler:2012:CCA:2094091.2094095}
Reinhard Tartler, Daniel Lohmann, Christian Dietrich, Christoph Egger, and
  Julio Sincero.
\newblock Configuration coverage in the analysis of large-scale system
  software.
\newblock {\em SIGOPS Oper. Syst. Rev.}, 45(3):10--14, January 2012.

\bibitem{TBKC:GPCE07}
Sahil Thaker, Don Batory, David Kitchin, and William Cook.
\newblock Safe composition of product lines.
\newblock In {\em Proc.\ Int'l Conf.\ Generative Programming and Component
  Engineering (GPCE)}, pages 95--104, New York, 2007. ACM Press.

\bibitem{TAKKSS:MD12}
Thomas Th{\"u}m, Sven Apel, Christian K{\"a}stner, Martin Kuhlemann, Ina
  Schaefer, and Gunter Saake.
\newblock Analysis strategies for software product lines.
\newblock Technical Report FIN-004-2012, School of Computer Science, University
  of Magdeburg, April 2012.

\bibitem{Thum:2014:CSA:2620784.2580950}
Thomas Th\"{u}m, Sven Apel, Christian K\"{a}stner, Ina Schaefer, and Gunter
  Saake.
\newblock A classification and survey of analysis strategies for software
  product lines.
\newblock {\em ACM Comput. Surv.}, 47(1):6:1--6:45, June 2014.

\bibitem{xterm}
{xterm}.
\newblock https://invisible-island.net/xterm/.

\bibitem{actstool}
Linbin Yu, Yu~Lei, Raghu~N Kacker, and D~Richard Kuhn.
\newblock Acts: A combinatorial test generation tool.
\newblock In {\em 2013 IEEE Sixth International Conference on Software Testing,
  Verification and Validation}, pages 370--375. IEEE, 2013.

\bibitem{Zave:2003:EFE:766951.766969}
Pamela Zave.
\newblock Programming methodology.
\newblock chapter An Experiment in Feature Engineering, pages 353--377.
  Springer-Verlag New York, Inc., New York, NY, USA, 2003.

\end{thebibliography}

\end{document}